\documentclass[prb,twocolumn,superscriptaddress,showpacs,preprintnumbers,amsmath,amssymb]{revtex4-2}
\usepackage{graphicx}
\usepackage{dcolumn}
\usepackage{bm}
\usepackage{tikz}
\usepackage{cancel}
\usepackage{tasks}
\usetikzlibrary{arrows,patterns,angles,quotes}
\usepackage[caption=false]{subfig}
\usepackage{float}
\usepackage{hyperref}
\hypersetup{
 bookmarks=true,   
 unicode=false,   
 pdftoolbar=true,  
 pdfmenubar=true,  
 pdffitwindow=false,  
 pdfstartview={FitH}, 
 pdfnewwindow=true,  
 colorlinks=true,  
 linkcolor=red,   
 citecolor=red,  
 filecolor=cyan,   
 urlcolor=magenta  
}

\newcommand{\nc}{\newcommand}
\nc{\half}{\frac{1}{2}}
\nc{\lamnr}{\lambda_{nr}}
\nc{\lamr}{\lambda_{r}}
\nc{\lamk}{\lambda_{k}}
\nc{\gp}{g({\bp})}
\nc{\gpp}{g({\bp'})}
\nc{\gpz}{g({\bp''})}
\nc{\gps}{g^{*}({\bp})}
\nc{\gpzs}{g^{*}({\bp''})}
\nc{\gpps}{g^{*}({\bp'})}
\nc{\bxi}{{\bf \xi}}
\nc{\bp}{{\bf p}}
\nc{\bpp}{{\bf p'}}
\nc{\bpz}{{\bf p''}}
\nc{\bk}{{\bf k}}
\nc{\bkp}{{\bf k'}}
\nc{\bkz}{{\bf k''}}
\nc{\bPi}{{\bf \Pi}}
\nc{\bera}{\langle}
\nc{\ket}{\rangle}
\nc{\bq}{{\bf q}}
\nc{\bqp}{{\bf q'}}
\nc{\tpi}{\tilde{\pi}}
\nc{\bpi}{\boldsymbol \pi}
\nc{\btpi}{\tilde{\boldsymbol \pi}}
\nc{\br}{{\bf r}}
\nc{\bfrac}[2]{\genfrac{}{}{0pt}{}{#1}{#2}}

\begin{document}

\title{Trion clustering structure and binding energy in 2D semiconductor materials: Faddeev equations approach}

\author{K.~Mohseni} 
\affiliation{%
Departamento de F\'isica, Instituto Tecnol\'ogico de Aeron\'autica,
DCTA, \\ 12228-900, S\~ao Jos\'e dos Campos, Brazil
}
\author{M.~R.~Hadizadeh} 
\email[Corresponding author: ]{mhadizadeh@centralstate.edu}
\affiliation{
College of Engineering, Science, Technology and Agriculture, Central State University, Wilberforce, OH 45384, USA
}
\affiliation{
Department of Physics and Astronomy, Ohio University, Athens, OH 45701, USA
}
\author{T.~Frederico} 
\affiliation{%
Departamento de F\'isica, Instituto Tecnol\'ogico de Aeron\'autica,
DCTA, \\ 12228-900, S\~ao Jos\'e dos Campos, Brazil
}
\author{D.~R.~da~Costa} 
\affiliation{
 Departamento de F\'isica, Universidade Federal do Cear\'a, Campus do Pici, 60455-900 Fortaleza, Cear\'a, Brazil
}
\author{A.~J.~Chaves} 
\affiliation{%
Departamento de F\'isica, Instituto Tecnol\'ogico de Aeron\'autica,
DCTA, \\ 12228-900, S\~ao Jos\'e dos Campos, Brazil
}

\date{\today}

\begin{abstract}
In this work, we develop the basic formalism to study trions in semiconductor layered materials using the Faddeev equations in momentum space for three different particles lying in two dimensions. We solve the trion Faddeev coupled integral equations for both short-range one-term separable Yamaguchi potential and Rytova-Keldysh (RK) interaction applied to the MoS$_2$ layer. 
We devise two distinct regularization methods to overcome the challenge posed by the repulsive electron-electron RK potential in the numerical solution of the Faddeev equations in momentum space. The first method regulates the repulsive interaction in the infrared region, while the second regulates it in the ultraviolet region.
By extrapolating the trion energy to the situation without screening, the two methods gave consistent results for the MoS$_2$ layer with a trion binding energy of $-49.5(1)$~meV for the exciton energy of $-753.3$~meV. We analyzed the trion structure for the RK and Yamaguchi potentials in detail, showing their overall similarities and the dominant cluster structure, where the strongly bound exciton is weakly bound to an electron. We found that this property is manifested in the dominance of two of the Faddeev components over the one where the hole is a spectator of the interacting electron pair.
\end{abstract}


\maketitle

\section{Introduction}
Few-body problems appear in physics at different scales, ranging from subatomic to celestial bodies. For semiconductors, the electron (e) and hole (h) can form bound states due to the electrostatic attraction. Speculated since the 30’s~\cite{Frenkel1931}, the exciton appears in the absorption spectrum of crystals, such as splitting of the lines in molecular crystals~\cite{David1964} or in band edge absorption features~\cite{gross1956optical}. Albeit the exciton is weakly bound due to the intrinsic screening of traditional semiconductors, the electron-hole interaction is fundamental to the understanding of the optical properties in semiconductors and insulators~\cite{knox1983}, as it follows from the works of G.~Dresselhaus~\cite{Dresselhaus1957} and Elliot~\cite{Elliott1957}. 

More complex few-body systems composed of holes and electrons were proposed by Lampert in 1958~\cite{Lampert1958}, such as the trion (eeh or ehh) and the biexciton. 
However, the weak binding energy of the trion, which results from the strong screening of the Coulomb interaction in ordinary materials, hindered its study until the advent of quantum wells. 
Although observed in 1977 in the asymmetric tail of exciton luminescence~\cite{Thomas1977}, a trion peak was only observed in 1993~\cite{Kheng1993} in quantum wells, effectively a two-dimensional (2D) system, whose energies were predicted to be an order of magnitude greater than in the three-dimensional (3D) case~\cite{Stebe1989} due to the quantum confinement effect. 

One interesting aspect of trion physics already noted by Lampert is the different limits as the hole and electron mass ratio changes, we can have the analog of H$_2^+$, H$^-$, and e$^-$e$^-$e$^+$. One should note that trions in semiconductors differ from traditional three-body systems such as the triton or the $^4$He$_3$ atomic trimer, as the constituents of the trion have two attractive interactions and one repulsive. We will explore this distinctive feature in this paper.

As we already mentioned, it was only with the dimension reduction that trions were first detected by observing the asymmetric tail of exciton luminescence~\cite{Thomas1977}. Three-particle bound states also appear in cold atom physics, where through trapping, the dimension can be reduced continuously from 3D to 2D~\cite{pethick2008bose}. The consequence is the disappearance of the Efimov effect and, together, the log-periodicity of the wave function, which turns into a power law~\cite{Rosa2022}. 

With the synthesis of 2D semiconductors~\cite{song2013two}, it was found that excitons can have huge binding energies~\cite{Ugeda2014} as also trions~\cite{Mak2013}, when compared to traditional materials. This happens due to the reduced screening, as the electric field lines lie outside the 2D semiconductor\cite{Chernikov2014}. In those systems, the strength of the interaction can be externally controlled by suitable dielectric engineering~\cite{Chaves2020}. Charge carriers in transition metal dichalcogenides (TMDs) interact mainly via the screened Coulomb interaction that in the classical regime is given by the Rytova-Keldysh potential, obtained as the solution of the Poisson equation for an infinitesimal thin dielectric slab~\cite{Cudazzo2011}.

Exciting prospects appear for few-body systems in novel 2D materials. There is a plethora of different materials that hosts excitons, trions, and biexcitons, such as TMDs, hexagon boron nitride, and graphene. In addition to that, excitons and trions can strongly couple with light, forming exciton-polaritons~\cite{Liu2015, Epstein_2020} and trion-polaritons~\cite{Emmanuele2020}, respectively. As the Fermi energy increases, e.g., with electrostatic doping, there is a transition of the trion to an exciton-Fermi polaron~\cite{Combescot2018}. The proximity effect~\cite{zutic2019}, which originates from short-range interactions, can also be used to tune the properties of excitons and trions through the suitable choice of van der Waals heterostructures, for example, the valley manipulation of excitons in TMDs due to the coupling with CrI$_3$~\cite{Seyler2018}, whose magnetization can be controlled by an external magnetic field, that breaks the time-reversal symmetry and the valley degeneracy.

Several experiments have already probed trions in 2D materials. Observations of large trion binding energies in MoS$_2$ reported experimental values between 20–43~meV for samples deposited on SiO$_2$ substrates~\cite{Mak_2012, Ross2013, Soklaski2014, Zhang2015} and 80~meV on suspended samples~\cite{Lin2019}, while Ref.~\cite{lin2014dielectric} measured for different substrates and found an extrapolation curve for the suspended case of 44~meV. Besides the dependence on the dielectric environment \cite{lin2014dielectric}, the trion binding energy depends on the doping \cite{Mak2013}, and also on the temperature \cite{PhysRevB.95.195427}. In this work (Sec.~\ref{sec:resultsRK}), we will discuss the trion in an undoped suspended MoS$_2$ layer at zero temperature, thus we do not expect an exact agreement with experimental measurements that are performed in a finite temperature and with residual doping.

There are already several theoretical calculations on trion binding energies~\cite{Berkelbach2013, PhysRevB.90.085419, PhysRevB.93.125423, Mayers2015, szyniszewski2017binding, Donck2017, Druppel_2017, kezerashvili2017trion, faddeev_configuration2018, Filikhin_2018, filikhin2018binding, Chang_2021, cavalcante2018stark}. The authors in Ref.~\cite{Donck2017} found a good agreement between the multiband and effective mass models, thus justifying our choice of using the effective mass approach in this work. In Ref.~\cite{Chang_2021}, it is reported a calculated binding energy for the trion in MoS$_2$ of 33.6~meV with a variationally optimized orbital approach ($m_e/m_0=$0.47, $m_h/m_0=$0.54 and $r=\,44.68$~\AA), in Ref.~\cite{PhysRevB.93.125423} the value of 33.7~meV was obtained by the stochastic variational method, and Ref.~\cite{cavalcante2018stark} reported the value of 32.1~meV by using an imaginary time evolution method for numerically solving the trion Schr\"odinger-like Hamiltonian. Based on the ab-initio many-body theory, a converged negatively charged intralayer trion binding energy was found to be of 58~meV \cite{Druppel_2017} with an exciton binding energy of $-0.76$~eV. The use of Faddeev equations in configuration space to calculate the charge positive and negative trion energies in various TMDs was reported in Refs. \cite{kezerashvili2017trion, faddeev_configuration2018, Filikhin_2018, filikhin2018binding}.

Our goal in this work is to study negatively charged trions within the Faddeev equations approach in momentum space and explore both the binding and structural properties of the trion in a MoS$_2$ layer. 
Within the adopted method, each Faddeev component is computed, which sums up the total wave function and carries information about each pair that composes the three-body state. The numerical convergence due to the repulsion between the electrons is a challenge, and to overcome this, we use two different approaches to regularize the electron-electron interaction at both long and short distances, to weaken the repulsion and turn the numerical calculations more accurate and finally, we extrapolate both results to compute the trion energy accurately. Furthermore, we cross-check the accuracy of our calculated trion energy by computing the expectation value of the Hamiltonian with the wave function. Our work addresses the following main points: (i) we provide a general discussion of the wave function properties for a 2D trion obtained within the Faddeev equations approach; (ii) a theoretical-numerical calculation of the trion binding energy in freestanding monolayer MoS$_2$ with different regularization schemes, with the accuracy checked by computing the expectation value of the Hamiltonian; and (iii) the degree of clusterization of the trion weakly bound state. 

The assumed theoretical framework is presented in Sec.~\ref{sec:Faddeev}, where we derive the Faddeev equations in 2D considering three different particles. In Sec.~\ref{sec:resultsYam}, we present results for the Yamaguchi model, a non-local separable and short-range potential~\cite{Yamaguchi1954, glockle2012quantum}, considering three attractive interactions and also for two attractive and one repulsive potential. In Sec.~\ref{sec:clusteryamaguchi}, we illustrate the cluster structure of the Yamaguchi model for trions. Sec.~\ref{sec:resultsRK} is devoted to presenting the results for the Rytova-Keldysh potential, where two different regularization procedures are introduced to compute the trion binding energy. In Sec.~\ref{sec:clusterRK}, it is illustrated the cluster structure of the trion by showing results for the total wave function and its Faddeev components, which we compare with the structure of the wave functions obtained by the two potential models. In Sec.~\ref{sec:summary}, we summarize the main findings of our study. This work is accompanied by six appendices where we detail our framework and numerical methods.

\section{\label{sec:Faddeev}
Faddeev equations for 3B bound states in two-dimension}
We consider the effective mass Hamiltonian for three-different particles
\begin{equation}
H=\sum_{i=1}^3\left(\frac{k_i^2}{2m_i}+V_i \right), \label{eq:hamiltonian}
\end{equation}
with $V_i\equiv V_i(r_j-r_k)$, $i\neq j\neq k$ and $m_i$ being the mass of the $i$-th particle. In the case of trions, this corresponds to the Wannier-Mott model.
The Schr\"{o}dinger equation for the bound state of three different particles interacting with pairwise interactions $V_i \equiv V_{jk}$ is given by
\begin{eqnarray}
\label{schrodinger_1}
\Psi =
\sum_{i=1}^{3}G_0 V_i\Psi = \sum_{i=1}^{3} \psi_i ,
\end{eqnarray}
where $\psi_i = G_0 V_i\Psi$ are the Faddeev components, $G_0=(E-H_0)^{-1}$ is the free propagator with three-body (3B) binding energy $E$ and free Hamiltonian $H_0$. Three Faddeev components $\psi_i$ satisfy the following coupled equations
\begin{equation}
\label{coupled_Faddeev_operator}
\psi_i = G_0 \,t_i \,(\psi_j+\psi_k\,) , 
\end{equation}
where $\{i,j,k\}$ is a cyclic permutation of $\{1,2,3\}$.
The two-body (2B) transition operators $t_i$ are defined by the Lippmann-Schwinger equation
\begin{equation}
\label{t_1}
t_i = V_i+V_i G_0 t_i.
\end{equation}
\begin{figure}[h!]
\includegraphics[width=3cm,angle=0]{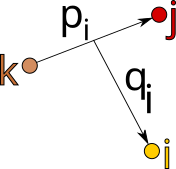}
\caption{Jacobi momenta convention used through the paper. $\{i,j,k\}$ correspond to the indices associated with the three particles, and $p_i$ and $q_i$ are their Jacobi momenta given in Eq.~\eqref{jacobi_lab}.}
\label{fig:schematics}
\end{figure}
To solve the coupled Faddeev equations \eqref{coupled_Faddeev_operator} in momentum space, we consider the 3B basis states $\vert \bp_i \bq_i \ket $, composed of two Jacobi momenta (see Fig.~\ref{fig:schematics}), which are defined in terms of the single particle momenta as
\begin{eqnarray}
\label{jacobi_lab}
\bp_i &=& \frac{m_k\bk_j-m_j\bk_k}{m_j+m_k} , \cr
\bq_i &=& \frac{m_j+m_k}{m_i+m_j+m_k}\left ( \bk_i-\frac{m_i}{m_j+m_k}(\bk_j+\bk_k)\right ) ,
\end{eqnarray}
where $\bp_i$ is the relative momentum of the pair $jk$, and $\bq_i$ is the relative momentum of the third particle $i$ with respect to the pair $jk$. The completeness relation of 3B basis states in 2D is defined as
\begin{eqnarray}
\label{completeness_relation}
\int d^2p_i \, \int d^2q_i \, \vert \bp_i \bq_i \ket \bera \bp_i \bq_i \vert 
= 1.
\end{eqnarray}
The projection of coupled Faddeev equations \eqref{coupled_Faddeev_operator} on 3B basis states $\vert \bp_i \bq_i \ket $ leads to three coupled 2D integral equations
\begin{eqnarray}
\label{Faddeev_cpupled_integral}
&&\psi_i(p_i,q_i,\phi_{i}) = \dfrac{1}{E_{3B}-\dfrac{p_i^2}{2\mu_{jk}} - \dfrac{q_i^2}{2\mu_{i,jk}}}\int_0^{\infty} dp'_i\,p'_i \int_{0}^{2\pi} d\phi'_i 
\cr 
&& \times \ t_i(p_i,p'_i, \phi'_i;\epsilon_i)
\Big[
 \psi_j(\mathcal{P}_{ji},\mathcal{Q}_{ji}, \phi_{ji} ) 
 +
 \psi_k(\mathcal{P}_{ki},\mathcal{Q}_{ki},\phi_{ki})
 \Big], \quad
\end{eqnarray}
where $\mu_{jk}= \dfrac{m_j m_k}{m_j+m_k}$ and $\mu_{i,jk} = \dfrac{m_i(m_j +m_k)}{m_i+m_j+m_k}$ are 2B and 3B reduced masses. The shifted momenta and angle quantities ${\mathcal P}_{ij}$, $\mathcal{Q}_{ij}$, and $\phi_{ij}$ are defined in Eq.~\eqref{eq:shifted}. The details of the derivation are given in Appendix~\ref{App_momentum_Faddeev}. 

The non-partial-wave 2B $t-$matrices $t_i(p_i,p'_i,\phi'_i ;\epsilon_i)$, with 2B subsystem energies $\epsilon_i= E_{3B}- \dfrac{q^2_i}{2\mu_{i,jk}}$, can be obtained from the summation of partial wave (PW) $t-$matrices $t_m(p_i,p'_i;\epsilon_i)$ as
\begin{eqnarray}
t_i(p_i,p'_i,\phi'_i ; \epsilon_i)&=& \frac{1}{2\pi} \,\sum^{\infty}_{m=0} \varepsilon_m \cos(m\phi'_i)\, t_m(p_i,p'_i ; \epsilon_i), 
\quad \\
&& \mbox{with}\quad\varepsilon_m= 
 \begin{cases}
 1 & m = 0 \\
 2 & m \neq 0\ \nonumber
 \end{cases}, 
 \label{Eq.t_pw_nonpw} 
\end{eqnarray} 
where PW projected 2B $t-$matrices in channel $m$, i.e., $t_m(p_i,p'_i,\epsilon_i)$, can be obtained from the solution of inhomogeneous Lippmann-Schwinger integral equation as
\begin{eqnarray}\label{eq:tmatrix}
t_m(p_i,p'_i; \epsilon_i)&=&V_m(p_i,p'_i)+\int_0^\infty dp''_i\, p''_i\, V_m(p_i,p''_i)\,\nonumber \\&\times&\frac{1}{\epsilon_i-\dfrac{p''^2_i}{2\mu_{jk}}}\,t_m(p''_i,p'_i ; \epsilon_i), \label{eq:LS}
\end{eqnarray}
with PW projected interactions obtained from
\begin{eqnarray}
V_m(p_i,p'_i) = \int_0^{2\pi} \, d\phi'_i \,V(p_i,p'_i,\phi'_i) \, \cos(m\phi'_i).
\end{eqnarray}
In the following sections, we present our numerical results for the solution of the coupled Faddeev equations ~\eqref{Faddeev_cpupled_integral} for trions with two different potentials: the short-range Yamaguchi potential and the long-range Rytova-Keldysh potential. The numerical solution details are provided in Appendix~\ref{sec:num}. 

\section{Trions: Yamaguchi potential}
\label{sec:resultsYam}

To test the formulation of the coupled Faddeev integral equations \eqref{Faddeev_cpupled_integral} and to validate our numerical solution, we first utilize the one-term separable potential with Yamaguchi-type form factors \cite{Yamaguchi1954,glockle2012quantum}
\begin{equation}
 V (p, p') = -\lambda g(p)g(p'), \quad g(p) = \frac{1}{(\beta^2 + p^2)^{m}},
\label{Yamagichi_pot}
\end{equation}
where the potential strength $\lambda$ can be obtained from the pole property of the 2B $t-$matrix at the 2B binding energy. We present our numerical results for 3B binding energies and wave functions using two different interaction combinations: (i) three attractive, and (ii) two attractive and one repulsive Yamaguchi-type potential, considering three particles with identical masses. 

We use three attractive Yamaguchi interactions to evaluate our formalism and computer codes for solving the general form of three coupled Faddeev integral equations in 2D. We should mention that this case is paradigmatic in cold-atom physics~\cite{blume2012few}, and it also appears in the formulation of the three-magnon bound state problem~\cite{nishida2013efimov}. In our context, these calculations  are valuable as  a preparation for the practical application involving trions. 

In order to clarify the assumed notation here, we denote the 3B binding energy as $E_{3B}$, being defined as the eigenvalue of the 3B Hamiltonian in Eq.~\eqref{eq:hamiltonian}, whereas the trion binding energy $E_\mathrm{t}$ is defined as the splitting between the 2B and 3B binding energies
\begin{equation}
E_\mathrm{t}=E_{3B}-E_{2B}.
\end{equation}
In Table~\ref{table_3B_Yamaguchi_attraction}, we present 3B binding energies obtained from the solution of the three coupled Faddeev integral equations \eqref{Faddeev_cpupled_integral} for three identical particles (mass = 1) interacting with three attractive Yamaguchi interactions. The input 2B $t-$matrices are obtained from the $s-$wave interactions. The calculated 3B and 2B binding energy ratios with different potential strengths $\lambda$ and form factor powers $m$ are in excellent agreement with the corresponding results from Ref.~\cite{adhikari1988efimov}. 

By solving the coupled Faddeev integral equations and having 3B binding energy and the Faddeev components, one can calculate the 3B wave function as a summation of three Faddeev components. In Appendix~\ref{App_3B_WF_momentum}, we show the details of the derivation of the 3B wave function in momentum space.
\begin{table}[t]
\centering
\caption{2B and 3B binding energies $E_{2B}$ and $E_{3B}$ calculated for three attractive Yamaguchi-type potentials with form factor parameter $\beta = 1$ and different powers $m$. The potential strength $\lambda$ is fitted to reproduce the desired 2B binding energy $E_{2B}$. The ratio of 3B and 2B binding energies $E_{3B}/E_{2B}$ are compared with corresponding results from Ref.~\cite{adhikari1988efimov}. The calculations are done with $\hbar c = \text{ mass} = 1$. }
\begin{tabular}{lcccccccccc}
\hline
 $\lambda$ && $E_{2B}$ && $E_{3B}/E_{2B}$ && $E_{3B}/E_{2B}$ \cite{adhikari1988efimov} \\
\hline
\multicolumn{7}{c}{$m=1$} \\
\hline
 $0.0602$ && $-0.0019$ && $ 9.21$ && $9.21 $ \\
 $0.0863 $ && $-0.0100$ && $ 6.83$ && $6.83 $ \\
 $0.1838$ && $-0.1000$ && $ 4.58$ && $4.58 $ \\
\hline
\multicolumn{7}{c}{$m=2$} \\
\hline
 $0.0801$ && $-0.0032$ && $ 7.30$ && $7.30 $ \\
 $0.1400$ && $-0.0211$ && $ 5.14$ && $5.14 $ \\
\hline
\multicolumn{7}{c}{$m=4$} \\
\hline
 $0.0481$ && $ -0.0001$ && $11.54$ && $11.53 $ \\
 $0.0731$ && $-0.0010$ && $7.91$ && $7.91 $ \\
 $0.1861$ && $-0.0200$ && $4.55$ && $4.55 $ \\
\hline
\multicolumn{7}{c}{$m=10$} \\
\hline
 $0.0561$ && $-0.0001$ && $10.05$ && $10.05 $ \\
 $0.0923$ && $-0.0010$ && $6.61$ && $6.61 $ \\
 $0.1562$ && $-0.0050$ && $4.89$ && $4.99 $ \\
\hline
\end{tabular}
\label{table_3B_Yamaguchi_attraction}
\end{table}
\begin{table}[thb]
\centering
\caption{Expectation values (EV) of 3B free Hamiltonian $\bera H_0\ket$, pair interactions $\bera V_i\ket$, total 2B interactions $\bera V \ket$, 3B Hamiltonian $\bera H \ket$, and eigenvalue ${E}_{3B}$ calculated for Yamaguchi-type potentials (three attractive (3A) in the second column, two attractive plus one repulsive interaction (2A+R) in the third column), given in Eq.~\eqref{Yamagichi_pot} with form factor parameters $\beta = m = 1$, and the potential strength $\lambda$ that reproduces dimer binding energy $E_{2B} = -0.1$. The relative percentage difference is Error=$ \vert ( \bera H \ket - {E}_{3B} ) / {E}_{3B} \vert \times 100 \%$.
The calculations are done with $\hbar c = \text{mass} = 1$.}
\begin{tabular}{ccc}
\hline
EV & 3A & 2A + R \\ \hline
 $\quad\bera H_0\ket\quad$ & $+0.46887$ & $+0.15756$\\ 
 $\bera V_1\ket$ & $-0.30904$ & $+0.03260$ \\ 
 $\bera V_2\ket$ & $-0.30904$ & $-0.15526$ \\
$\bera V_3\ket$ & $-0.30904$ & $-0.15523$ \\
$\bera V \ket$ & $-0.92712$ & $-0.27789$ \\
$\bera H \ket$ & $-0.45825$ & $-0.12033$ \\
${E}_{3B}$ & $-0.45824$ & $-0.12034$ \\
Error & $+0.00218$ & $+0.00831$ \\
 \hline
\end{tabular}
\label{table_Yamaguchi_attractions_expectation}
\end{table}
To test the accuracy of the 3B wave function in momentum space, in Table~\ref{table_Yamaguchi_attractions_expectation}, we compare the expectation values of 3B Hamiltonian with the calculated 3B binding energy for the factor parameters $\beta = m = 1$ and the potential strength $\lambda$ that reproduces 2B binding energy $E_{2B} = -0.1$. The separable potential strength is obtained by introducing Eq.~\eqref{Yamagichi_pot} in Eq.~\eqref{eq:LS}, and considering that the t-matrix has a pole at $\epsilon_i=E_{2B}$, then the potential strength can be obtained by
\begin{equation}
\lambda^{-1}=- 2\pi \int_0^\infty dp_i^{\prime\prime} p_i^{\prime\prime}\frac{|g(p^{\prime\prime})|^2}{E_\mathrm{2B}-\frac{{p_i^{\prime\prime}}^2}{2\mu_{jk}}}.
\end{equation}
The expectation values of the kinetic energy and potential in the exciton state are, in this case, given by
\begin{equation}\label{eq:expectationvaluesexcitonYama}
 \langle H_0\rangle= 0.138407\quad \text{and}\quad 
\langle V\rangle= -0.238407 .
\end{equation}

As we can see in Table~\ref{table_Yamaguchi_attractions_expectation}, the 3B binding energy and the expectation value of Hamiltonian are in excellent agreement. 
The details of the calculation of expectation values of Hamiltonian $\bera H\ket$ from the expectation value of 3B free Hamiltonian $\bera H_0\ket$ and 2B interactions $\bera V_i \ket$ are given in Appendix~\ref{App_expectations}.

Some interesting qualitative aspects can be seen in Table~\ref{table_Yamaguchi_attractions_expectation}. When the sign of the potential $V_1$ is changed and becomes repulsive, the state swells due to the dramatic decrease in the splitting of the 2B and 3B energies, namely from $|E_{3B}-E_{2B}| = 0.3582$ to $0.0203$. Consequently, the kinetic energy is also reduced to about one-third of the value obtained with only attractive potentials. Due to the repulsion, the wave function is depleted when the relative distance between particles 2 and 3 lies in the range of the potential, and the expectation value $\langle V_1\rangle$ turns to be negative and reduced to one-tenth with respect to the attractive case. Furthermore, the expectation values of $\langle V_2\rangle$ and $\langle V_3\rangle$ are also halved, and due to our choice of mesh points, the equality $\langle V_2\rangle=\langle V_3\rangle$ is fulfilled to 0.02\%, which is reflected in error around 0.008\% in the computation of $\langle H\rangle$, which is four times larger than the error in the attractive case.
\begin{figure}[thb]
\centering
\begin{tabular}{cc}
\includegraphics[width=4.5cm]{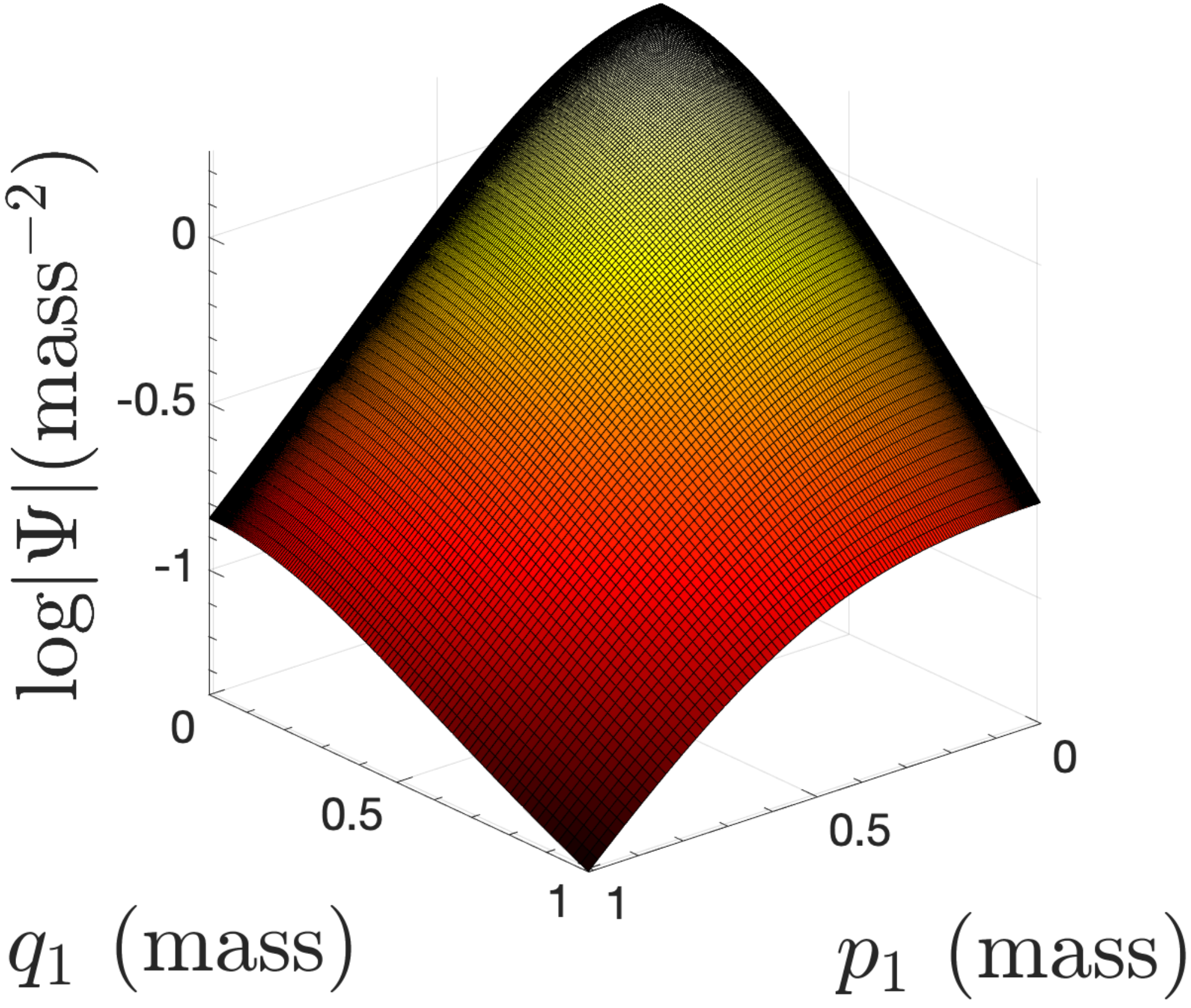} 
\includegraphics[width=4.2cm]{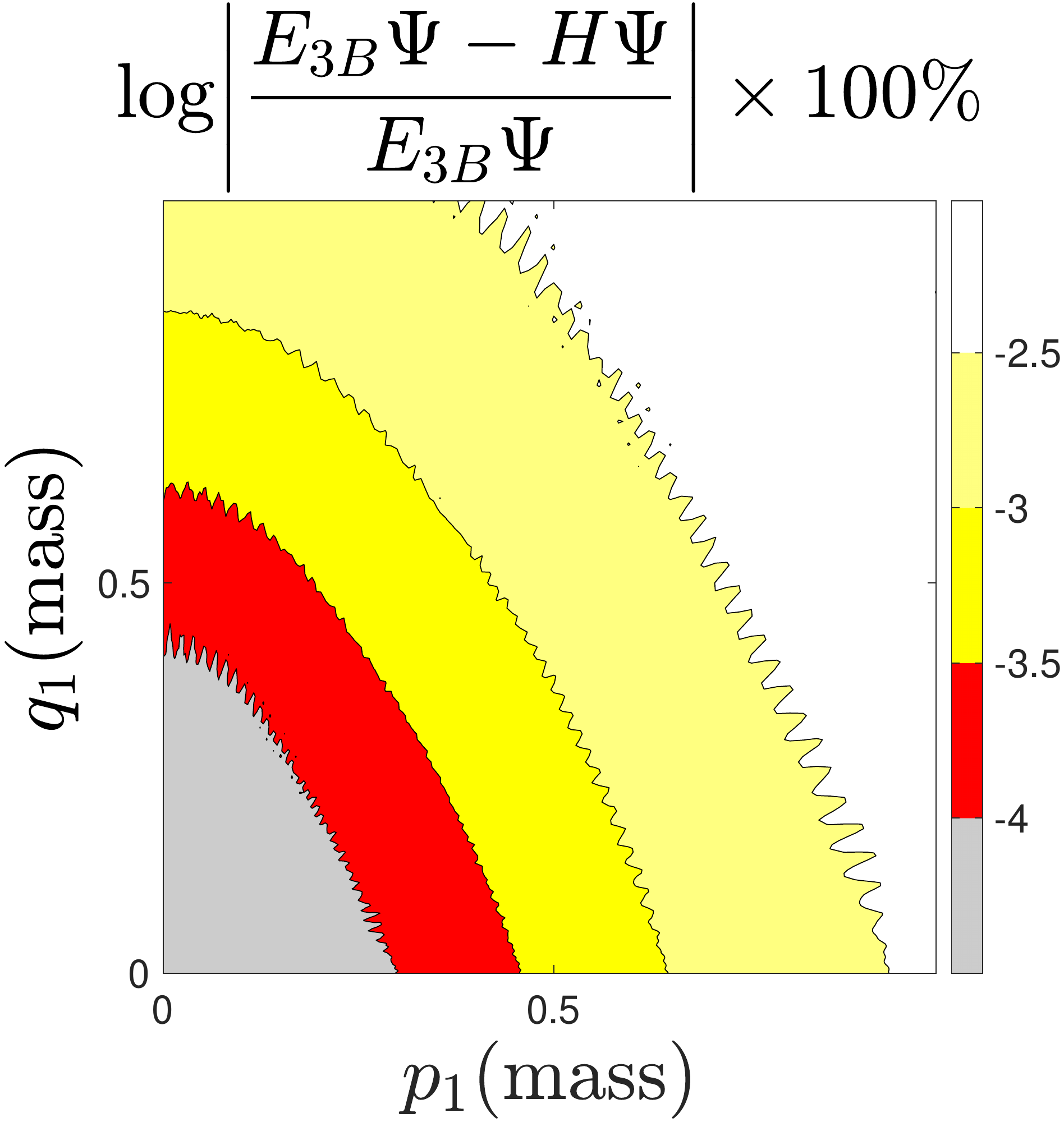}
\end{tabular}
\caption{Left panel: 3D plot of the 3B wave function for three attractive Yamaguchi-type potentials (3A). Right panel: relative error. All plots are for the angle $\phi_1 = 0$. The results are obtained with form factor parameters $\beta = m = 1$, and the potential strength $\lambda$ that reproduces 2B binding energy $E_{2B} = -0.1$. The calculations are done with $\hbar c = \text{mass} = 1$.}
\label{WF1_all_attraction}
\end{figure}
\begin{figure}[t]
\centering
\begin{tabular}{cc}
\includegraphics[width=4.5cm]{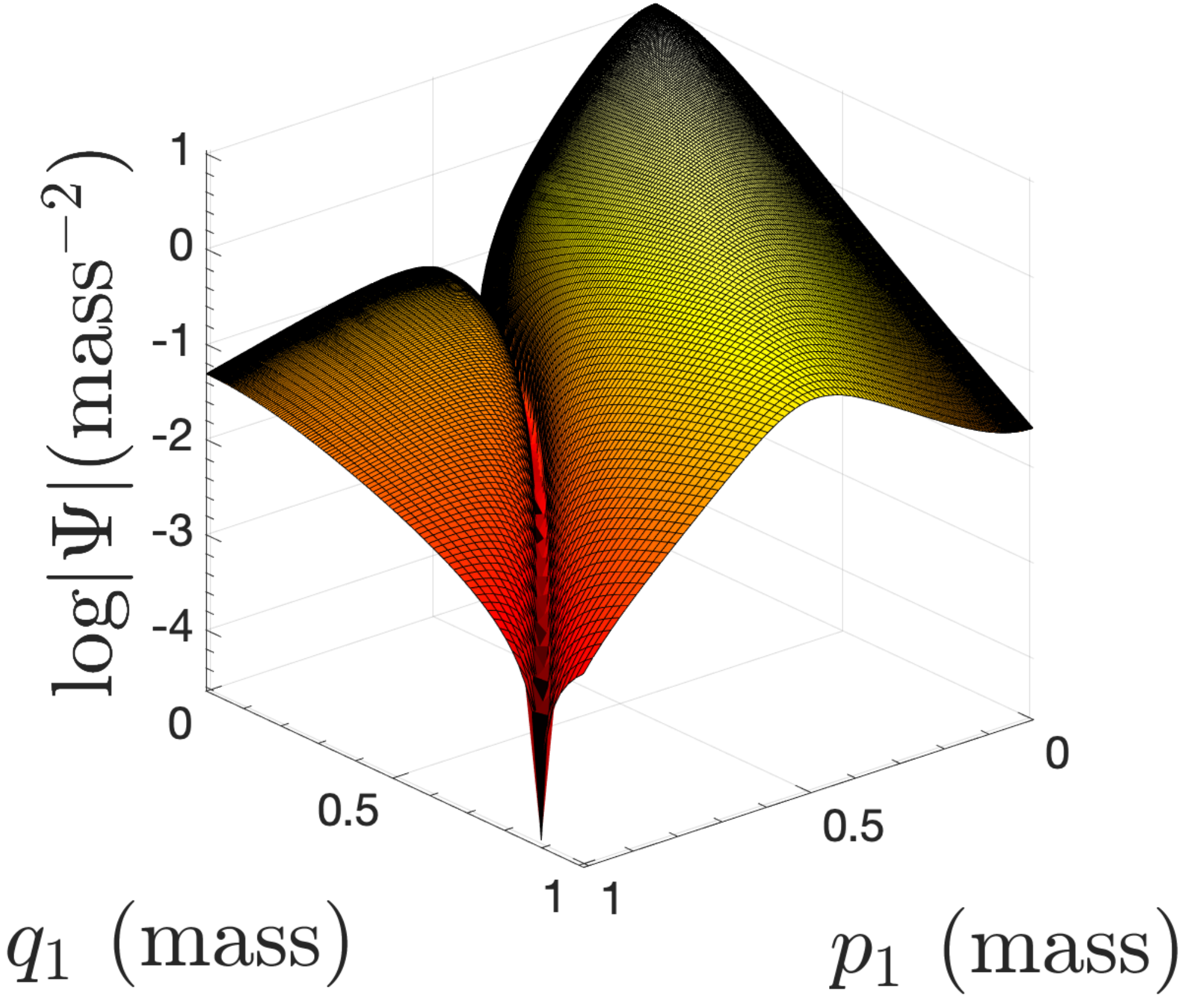} &
\includegraphics[width=4.2cm]{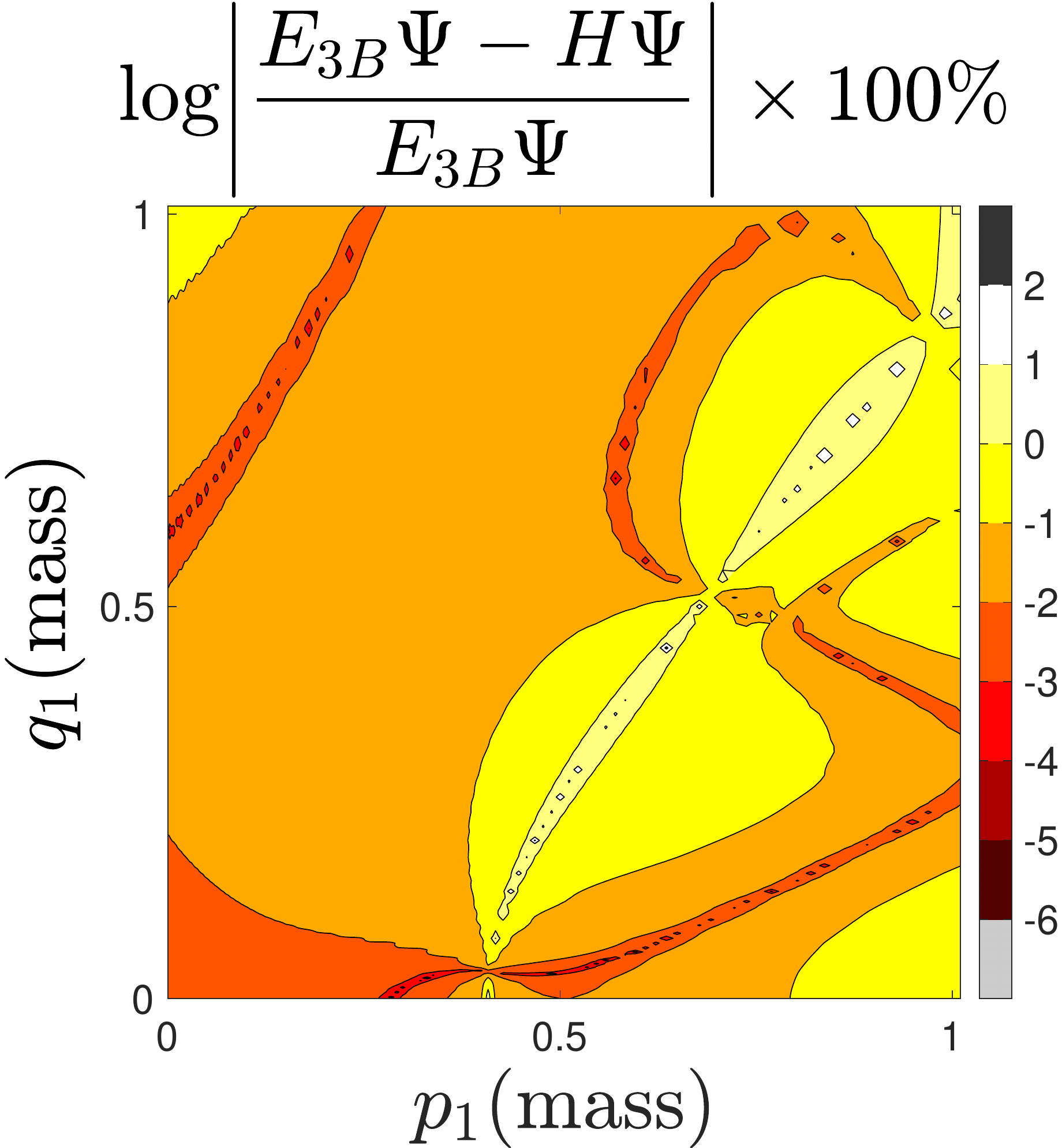}
\end{tabular}
\caption{Left panel: 3D plot of the 3B wave function for one repulsive $(V_1)$ and two attractive Yamaguchi-type potentials (2A + R). Right panel: relative error.
All plots are for the angle $\phi_1 = 0$. The results are obtained with form factor parameters $\beta = m = 1$, and the potential strength $\lambda$ that reproduces 2B binding energy $E_{2B} = -0.1$. The calculations are done with $\hbar c = \text{mass} = 1$.}
\label{WF1_2A_R}
\end{figure}

The cluster structure of the trion is indicated by its kinetic and potential energies when comparing the results of $\bera H_0\ket$, $\bera V_2\ket$ and $\bera V_3\ket$ from Table~\ref{table_Yamaguchi_attractions_expectation}, with the expectation values of the kinetic and potential energies of the exciton given in Eq.~\eqref{eq:expectationvaluesexcitonYama}. The screening of the hole that composes the exciton in the trion~\cite{Filikhin_2018} weakens the electron-exciton interaction favoring the formation of the remarkable cluster structure. Note that the electrons should be in a singlet spin state or an antisymmetric combination of different valley states.

\section{Yamaguchi Trion Clusterization}
\label{sec:clusteryamaguchi}

In Figs.~\ref{WF1_all_attraction} and \ref{WF1_2A_R}, we show the magnitude of the 3B wave function obtained in two cases with Yamaguchi interactions as a function of the magnitude of Jacobi momenta $p_1$ and $q_1$ for the angle between them $\phi_1 = 0$. The first case shown in Fig.~\ref{WF1_all_attraction} is the eigenstate of the 3B system with three attractive potentials (3A), and the second case shown in Fig.~\ref{WF1_2A_R} is the wave function for the weakly bound state obtained with one repulsive potential, $V_1$, and two attractive ones (2A+R). We also show the relative percentage error for the verification of the Schr\"odinger equation, defined in Appendix~\ref{App_test_Schrodinger}, with 3B wave function and binding energy.

The plot of the wave function for the 3A case for $\bp_1$ and $\bq_1$ aligned presented in the left panel of Fig.~\ref{WF1_all_attraction} shows that the momentum distribution is somewhat symmetric due to the identical masses and the bosonic symmetry of the system. When the repulsive potential $V_1$ is introduced in the case 2A+R, seen in the left panel of Fig.~\ref{WF1_2A_R}, the wave function develops a node line as well becomes more sharply peaked around the origin, the latter due to the small binding energy (see Table~\ref{table_Yamaguchi_attractions_expectation}). The numerical accuracy of our calculations is checked through the ratio $|(E\Psi-H\Psi)/(E\Psi)|$ and presented in the right panels of Figs.~\ref{WF1_all_attraction} and \ref{WF1_2A_R} for the 3A and 2A+R cases, respectively. As expected, the results for the 3A case show quite good numerical accuracy, while the 2A+R results, mainly outside the node, are also accurate. As expected, the region of the largest errors for the 2A+R case in the right panel of Fig.~\ref{WF1_2A_R} follows the node of the 3B wave function.
\begin{figure}[thb]
\centering
\begin{tabular}{cc}
\includegraphics[width=4.2cm]{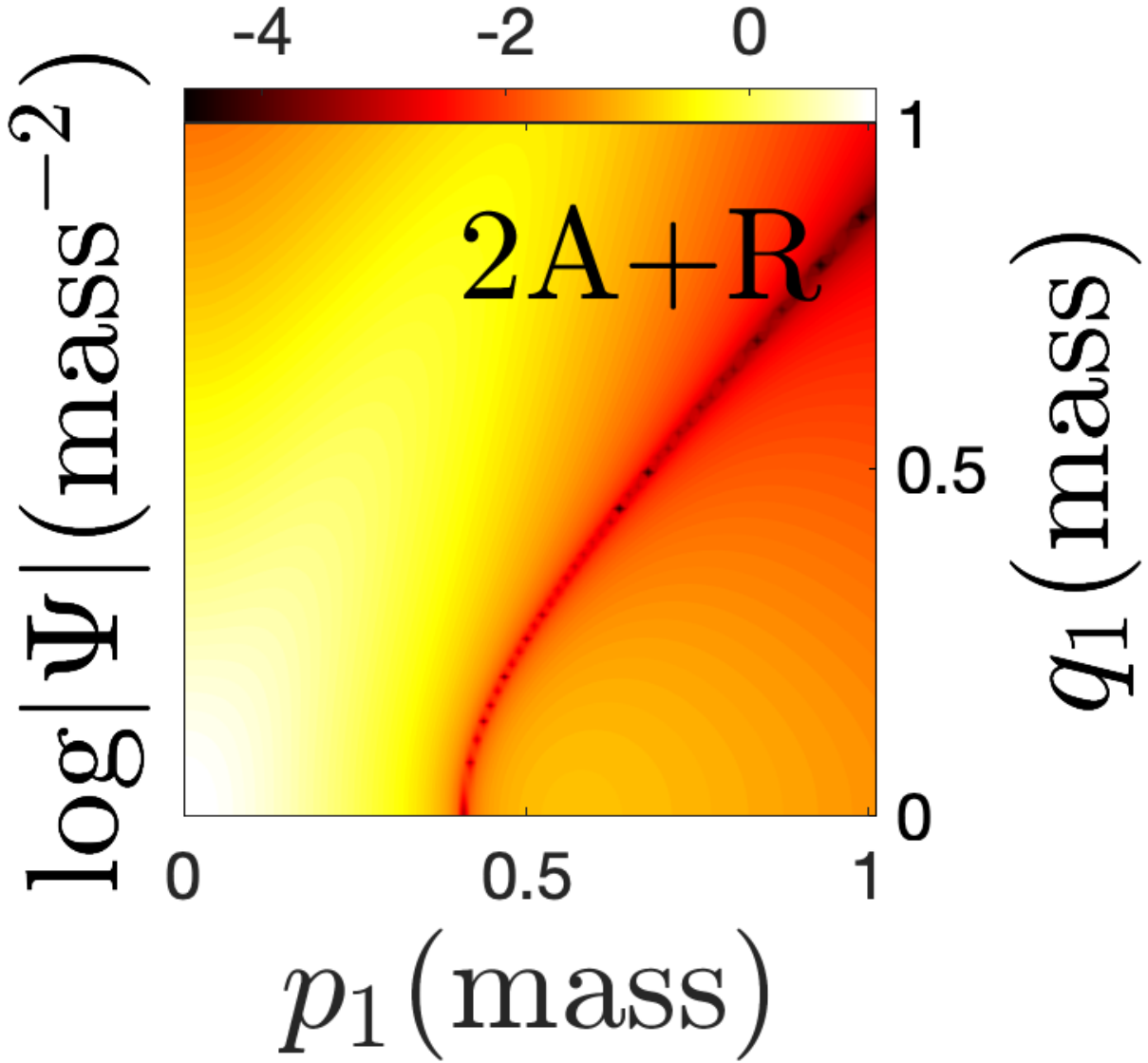}
&
\includegraphics[width=4.2cm]{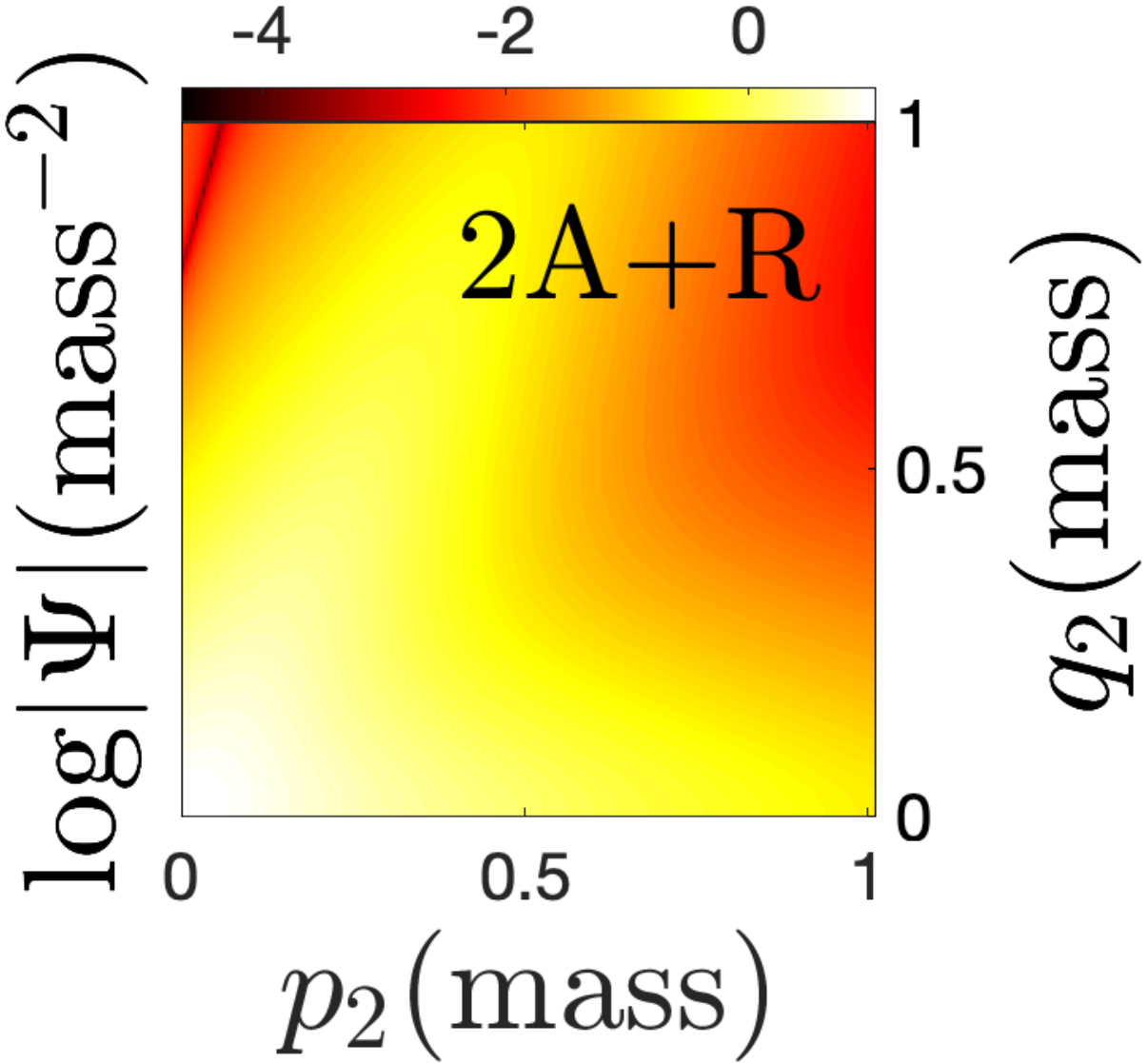}
\end{tabular}
\caption{Contour plot of the 3B wave function calculated for three Yamaguchi-type potentials as a function of the magnitude of Jacobi momenta.
$V_1$ repulsive, $V_2$ and $V_3$ attractive. The plots are for the angle $\phi_1 = \phi_2 = 0$. The results are obtained with form factor parameters $\beta = m = 1$ and the potential strength $\lambda$ that reproduces 2B binding energy $E_{2B} = -0.1$. The calculations are done with $\hbar c = \text{mass} = 1$.}
\label{WF1_AR}
\end{figure}
In Fig.~\ref{WF1_AR}, we present the contour plots of the 2A+R model in the ($p_1\times q_1$)--plane for $\phi_1=0$ (left panel) and ($p_2\times q_2$)--plane with $\phi_2=0$ (right panel). The node line, visible in the left panel of Fig.~\ref{WF1_AR}, similarly to the left panel in Fig.~\ref{WF1_2A_R}, comes from the cancellation between $\psi_1$ and $\psi_2+\psi_3$ due to the reversed sign of $\psi_1$ with respect to $\psi_2$ and $\psi_3$ from the repulsive potential $V_1$, then
\begin{equation}
 \psi_1(p_1,q_1,0)=-\psi_2(p_2,q_2,\phi_2)-\psi_3(p_3,q_3,\phi_3)\, ,
\end{equation}
where $\phi_{2,3}=0$, and these relations implicitly define the node line, understood by rewriting the momenta labeled by 2 and 3, according to
\begin{subequations}
\begin{eqnarray}\label{eq:q2p2q3p3}
&&\bp_2 =-\frac12 \bp_1 -\frac34 \bq_1 ,\quad\bq_2 = \bp_1-\frac12\bq_1,\\
&&\bp_3 =-\frac12 \bp_1 +\frac34 \bq_1 ,\quad\bq_3 = -\bp_1-\frac12\bq_1.
\end{eqnarray}
\end{subequations}
The node line is barely seen in the right panel of Fig.~\ref{WF1_AR}, with the momenta expressed in terms of $p_2$ and $q_2$ with $\phi_2=0$. 

Another property of the wave function is the well-defined maximum seen in both Figs.~\ref{WF1_2A_R} and \ref{WF1_AR}, which can be qualitatively understood by a semi-classical picture and the prevalence of the cluster structure. This dominant configuration suggests that the electron and hole (denoted as particle 1) are very close and ``moving together'', which relates the momenta $p_1$ and $q_1$ and provides an interpretation of the pattern of the maximum found in the left panel of Fig.~\ref{WF1_2A_R} and around the whitish-yellow color of Fig.~\ref{WF1_AR}, namely, along the line $p_1\propto q_1$ for the 2A+R model. In other words, the relative velocity between the two electrons is the same as the one between the far-apart electron and the hole, which forms the strongly bound exciton, which is very clear in the situation where $\phi_1=0$. This also explains the obtained pattern of the maximum of the wave function, with two branches observed in the right panel of Fig.~\ref{WF1_2A_R} in the ($p_2\times q_2$)--plane, as we shall discuss in more detail in what follows. We would like to draw the reader's attention to the practical significance of such plots, which provide insights into the regions where the wave function is more substantial. This information is crucial for distributing mesh points appropriately to obtain precise solutions to the Faddeev equations.
\begin{table}[t]
\caption{The inner product of the Faddeev components $\bera \psi_i | \psi_j \ket $ and their contribution to the normalization of the 3B wave function $|\Psi \ket$.}
 \begin{tabular}{cccc}
 \hline
 & $j=1$ & $j=2$ & $j=3$ \\ 
 \hline
 $i=1$ & $0.0560$ & $-0.1047$ & $-0.1047$ \\ 
 \hline
 $i=2$ & $-0.1047$ & $0.4400$ & $0.2413$ \\
 \hline
 $i=3$ & $-0.1047$ & $0.2413$ & $0.4403$ \\
 \hline
 \end{tabular}\label{tab:Yamoverlaps}
\end{table}
We would like to emphasize that each Faddeev component of the wave function in our system carries the asymptotic form of the total wave function in each pairwise interaction channel~\cite{faddeev1960scattering}. Specifically, in our context, the Faddeev component $\psi_1$ at asymptotically large distances of the hole to the center of mass of the electron-electron interacting pair decays exponentially, indicating that the two electrons are in a continuum state. Similarly, at asymptotically large distances of the spectator electron (particle 2) to the center of mass of the electron-hole system (particles 3 and 1), the Faddeev component $\psi_2$ decays exponentially, signifying that this pair necessarily forms the strongly bound exciton state. The same reasoning applies to $\psi_3$, where the electron-hole pair is formed by particles 1 and 2.

A scheme illustrates the clustering of the wave function: $\psi_2\sim$~[3(e)1(h)]---2(e) and $\psi_3\sim$~[1(h)2(e)]---3(e), which should be the two dominant configurations, with the electrons in the spin singlet state or an antisymmetric combination of different valley states. Indeed in Table~\ref{tab:Yamoverlaps}, one observes that the Faddeev component $\psi_1$ is suppressed with respect to $\psi_2$ and $\psi_3$ by one order of magnitude considering the inner products. With that in mind, we should now look to Fig.~\ref{WF1_AR} (right panel) for the modulus of the total wave function, $|\Psi|$ in the ($p_2\times q_2$)--plane. We identify two branches where $|\Psi|$ is larger: one for small $q_2$ and a diagonal one. The lower branch corresponds to the contribution of $\psi_2$ for $q_2\approx 0$, which is the relative momentum of the weakly bound spectator particle 2(e) with respect to the strongly correlated pair of particles 1 and 3.
The spread in the values of $p_2$ is associated with the small size of the strongly bound exciton in the [3(e)1(h)]---2(e) configuration. The diagonal branch, $p_2\propto q_2$, where the momentum probability density is enhanced, corresponds to the dominance of $\psi_3$ associated with the [1(h)2(e)]---3(e) configuration. In this case, electron 2 moves together with hole 1, as the exciton is strongly bound, and electron 3 is the spectator. 

\section{Trions: Rytova-Keldysh potential}
\label{sec:resultsRK}
 
\begin{figure}[b] 
\centering 
 \includegraphics[width=1\linewidth]{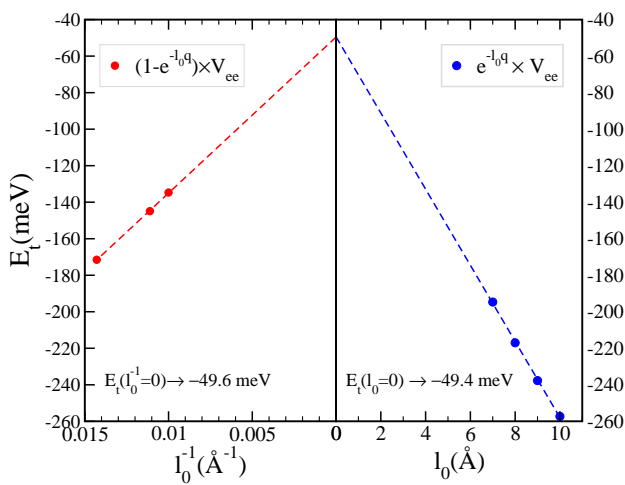}
\caption{Extraction of trion binding energy $E_t$ with a linear extrapolation on energies obtained from two screenings (see Table~\ref{v1_v2_screening}) at the physical points $l^{-1}_0=0$ \AA$^{-1}$ (left panel) and $l_0=0$ \AA (right panel).}
\label{fig_trion_extrapolations}
\end{figure}

Building on our understanding gained from the 2A+R Yamaguchi potential model in 2D, we now study the trion binding energy and structure for the MoS$_2$ layer with the Rytova-Keldysh potential.
The Rytova-Keldysh electron-hole (e-h) [electron-electron (e-e)] interaction in momentum space is given by~\cite{rytova}
\begin{equation}
\label{eh_RK}
V_{\bfrac{eh}{ee}}(q) = \pm \frac{1}{4\pi^2} \left( \frac{1}{4\pi \epsilon_0} \frac{2\pi e^2}{q(1 + r_0 q )}\right) ,
\end{equation}
where the momentum transfer is defined by $\vert \bq \vert = \vert \bp-\bp' \vert$. The parameters of the e-e and e-h potentials for the MoS$_2$ layer are given in Table~\ref{table_RK_parameters}. The value of the screening length $r_0$ is fitted to give an exciton binding energy of $-753$~meV in agreement with the value obtained from the measurement of the exciton position in the absorption spectrum of a suspended MoS$_2$ layer~\cite{klots2014probing} and the corresponding GW bandgap~\cite{Zhang_2016}. For our reference, the expectation values of the kinetic and potential energies in the exciton state are
\begin{equation}
\langle H_0\rangle=214.64\,\text{meV}\quad\text{and}\quad \langle V\rangle=-967.96\,\text{meV} \, ,
\end{equation}
which are related to the manifestation of this strongly bound two-particle system.
\begin{table}[t]
\centering
\caption{The used parameters in our calculations for the Rytova-Keldysh electron-hole and electron-electron interactions, defined in Eq.~\eqref{eh_RK}, for monolayer MoS$_2$. }
\begin{tabular}{c|ccc} 
\hline
$r_0$  & $27.05$ \AA \\ 
$\epsilon_0 / e^2$ & $\dfrac{1}{4\pi \alpha} \cdot \dfrac{1}{\hbar c}$ K$^{-1} \cdot$ \AA$^{-1}$ \\
$\alpha$ & $137.035999084$ \\
$m_e$ & $0.47 \ m_0$ \cite{Korm_nyos_2015}\\ 
$m_h$ & $0.54 \ m_0$ \cite{Korm_nyos_2015} \\ 
$m_0$ & $0.510998950$ MeV \\
1 eV & $1.160451812 \cdot 10^4$ K \\
$\hbar c$ & $1973.269804 $ eV$\cdot$ \AA \\ 
\hline
\end{tabular}\label{table_RK_parameters}
\end{table}
The example studied in Sec.~\ref{sec:clusteryamaguchi} has already taught us that the accuracy of our numerical solution of the Faddeev equations decreases in the case of 2A+R Yamaguchi potential with respect to the 3A attractive case (cf. Table~\ref{table_Yamaguchi_attractions_expectation}). This expected behavior of our numerical solutions is due to the small trion binding energy and the node in the wave function. On top of that, considering that the Rytova-Keldysh potential is of a longer range when contrasted to the Yamaguchi model, the numerical solution becomes more challenging due to the competition between attraction and repulsion with the same strength. To make this issue numerically amenable, the repulsive Rytova-Keldysh potential between the electrons is screened by two different regulators \cite{deltuva2005momentum}, namely
\begin{equation}
 V(q)\to ( 1 - e^{-l_0 q} ) V_{ee}(q)\quad\text{ or}\quad e^{-l_0\,q} V_{ee}(q)\,,
 \label{eq:RKregulated}
\end{equation}
where in the first case, the Rytova-Keldysh potential is damped at small momentum or large distance, while the second one is at large momentum or small distance.
\begin{figure}[t]
\centering 
\includegraphics[width=1\linewidth]{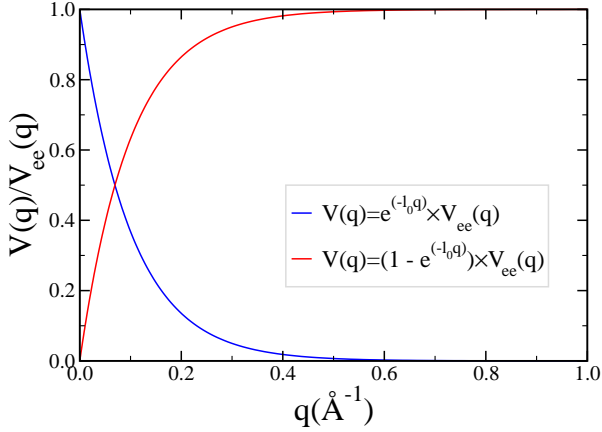}
\caption{The screening function $V(q)/V_{ee}(q)$ as a function of the momentum transfer $q$ for (blue curve) $V(q) \to e^{-l_0\,q} V_{ee}(q)$ and (red curve) $V(q) \to ( 1 - e^{-l_0 q} ) V_{ee}(q)$ with screening parameter $l_0=10$ \AA.}
\label{fig_RK_screening}
\end{figure}
\begin{table}[b]
\centering
\caption{Expectation values (EV) in meV of the 3B free Hamiltonian $\bera H_0\ket$, pair interactions $\bera V_i\ket$, total 2B interactions $\bera V \ket$, 3B Hamiltonian $\bera H \ket$, and binding energy $E_{3B}$ in meV calculated for Rytova-Keldysh potentials (two attractive plus one repulsive interaction (2A+R)) given in Eq.~\eqref{eh_RK} with screening parameter $l_0 =100$~\AA ~in the scheme $V_1(q)\to ( 1 - e^{-l_0 q} ) V_{ee}(q)$. The relative percentage difference is Error=$ \vert ( \bera H \ket - E_{3B} ) / E_{3B} \vert \times 100 \%$.}
\begin{tabular}{cc}
\hline
EV & 2A + R \\ \hline
 $\quad\bera H_0\ket\quad$ & $+247.66$\\ 
 $\bera V_1\ket$ & $+443.39$ \\ 
 $\bera V_2\ket$ & $-825.90$ \\
$\bera V_3\ket$ & $-825.76$ \\
$\bera V \ket$ & $-1208.27$ \\
$\bera H \ket$ & $-960.61$ \\
$E_{3B}$ & $-960.58$ \\
Error & $+0.00312$ \\
 \hline
\end{tabular}
\label{table_RK_att_rep_expectation}
\end{table}
Fig.~\ref{fig_RK_screening} illustrates quantitatively both screenings [Eq.~\eqref{eq:RKregulated}] with $l_0=10$~\AA. In our actual calculations, the results for the trion binding energy will be obtained by performing the extrapolation to $l_0=0$. Before that, the results of the expectation values for the Rytova-Keldysh potential screened at low momenta for $l_0=100$~\AA ~are depicted in Table~\ref{table_RK_att_rep_expectation}. Due to the contribution of the spectator electron external to the exciton, the expectation value of the kinetic energy is somewhat larger than the one found for the exciton given in Eq.~\eqref{eh_RK} with a value of 214.64~meV compared to the trion one of 247.66~meV. The expectation values of the attractive potentials $V_2$ and $V_3$ are somewhat less in magnitude than the one for the exciton of -967.96~meV. In the trion magnitude of the potential energy of the repulsive potential is about one-half of the attractive one. This last feature can be understood as the electrons should be more separated than the relative distance within the strongly bound electron-hole pair. While the trion and exciton splitting is 207.29~meV, it shows a weakly bound trion with respect to the exciton. Table~\ref{table_RK_att_rep_expectation} also indicates a good accuracy found in our solution by comparing the results from the expectation value of the Hamiltonian $\bera H \ket$ and the energy $E_{3B}$ obtained by solving the coupled Faddeev integral equations, which shows a deviation of only 0.003\% between these two values. 
\begin{table}[t]
\centering
\caption{Trion ground state binding energies ($E_{3B}$) for different screening parameter $l_0$ obtained from two screening schemes shown in Fig.~\ref{fig_RK_screening} and given in Eq.~\eqref{eq:RKregulated}.}
\begin{tabular}{c|c||c|c}
 \multicolumn{2}{c||}{$V(q) \to e^{-l_0\,q} V_{ee}(q)$} &\multicolumn{2}{c}{$V(q) \to ( 1 - e^{-l_0 q} ) V_{ee}(q)$} \\
 \hline
$l_0$ (\AA) & $E_{3B}$ (meV) & $l_0$ (\AA) & $E_{3B}$ (meV) \\
 \hline
$25$  & $-1195.1$ & $1$ & $-1444.8$ \\
\hline
$20$  & $-1150.6$ & $5$ & $-1309.6
$ \\
\hline
$17$  & $-1117.3$ & $10$ & $-1212.0$ \\
\hline
$15$  & $-1091.5$ & $15$ & $-1147.7$ \\
\hline
$13$  & $-1062.2$ & $20$ & $-1101.2$ \\
\hline
$11$  & $-1028.9$ & $30$ & $-1037.3$ \\
\hline
$10$  & $-1010.6$ & $50$ & $-965.6$ \\
\hline
$9$  & $-991.0$ & $70$ & $-924.8$ \\
\hline
$8$  & $-970.3$ & $90$ & $-898.2$ \\
\hline
$7$  & $-948.0$ & $100$ & $-888.0$ \\
\hline \end{tabular}
\label{v1_v2_screening}
\end{table}
In Appendix~\ref{App_trion_energies}, we present a convergence study of the trion energy, as summarized in Table~\ref{table_trion_l0}, which requires the extrapolation in the number of quadrature points. Noteworthy that the results presented in Table~\ref{table_RK_att_rep_expectation} are not converged in terms of the number o quadrature points but are good enough to compute the expectation value of the Hamiltonian, which should be interpreted as a lower bound. The extrapolated results from Table~\ref{table_trion_l0} are collected in Table~\ref{v1_v2_screening} for the two forms of the screening implemented for the Rytova-Keldysh electron-electron repulsive potential. Fig.~\ref{fig_trion_extrapolations} shows this extrapolation as a function of (left panel) $l_0^ {-1}$, for the short distance screening trion energies of the repulsive potential and (right panel) $l_0$, for the large distance screening. As shown in Fig.~\ref{fig_trion_extrapolations}, the results exhibit a perfect linear behavior which allows an accurate extrapolation to the trion binding energy. The linear extrapolation on binding energies obtained from the first screening (left panel) on the domain $[70,100]$ \AA$^{-1}$ leads to a trion binding energy of $-49.6$ meV, while a linear extrapolation on the second screening (right panel) on the domain $[7,10] $ \AA~ leads to a trion binding energy of $-49.4$ meV. These results lie in the range of previous experiments reported in Refs.~\cite{lin2014dielectric, Lin2019}.
\begin{table}[b]
 \caption{The inner product of the Faddeev components $\bera \psi_i | \psi_j \ket $ and their contributions in the normalization of the 3B wave function $|\Psi \ket$ obtained for the screening parameter $l_0 =100$~\AA~in the screening scheme $V_1(q)\to ( 1 - e^{-l_0 q} ) V_{ee}(q)$.}
 \begin{tabular}{cccc}
 \hline
 & $j=1$ & $j=2$ & $j=3$ \\ 
 \hline
 $i=1$ & $0.1772$ & $-0.2808$ & $-0.2806$ \\ 
 \hline
 $i=2$ & $-0.2808$ & $0.5197$ & $0.4534$ \\
 \hline
 $i=3$ & $-0.2806$ & $0.4534$ & $0.5190$ \\
 \hline
 \end{tabular}
 \label{tab:FaddeevRK}
\end{table}
\begin{figure*}[thb] 
\centering 
\includegraphics[width=.9\linewidth]{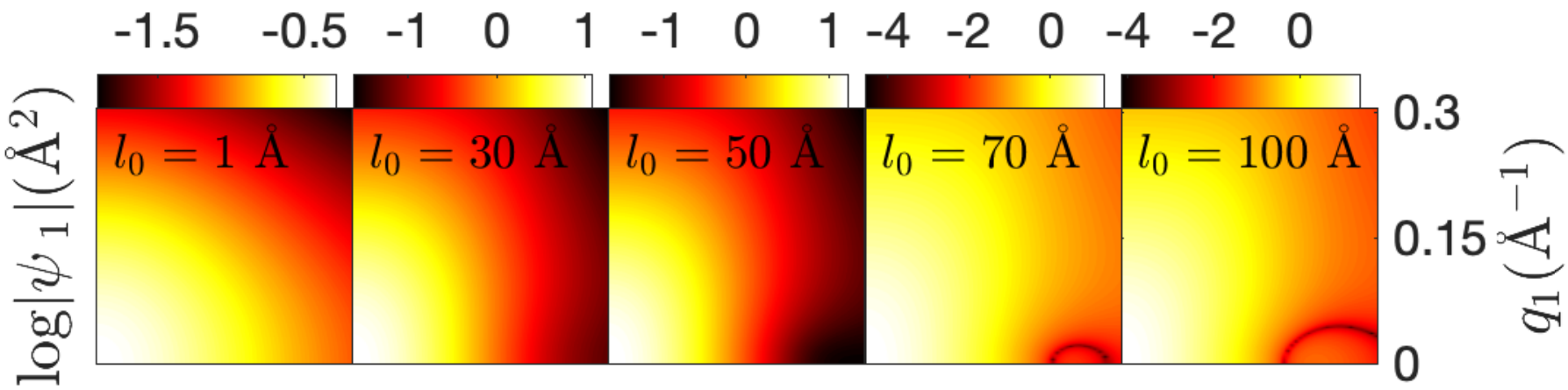}
\\
\includegraphics[width=.9\linewidth]{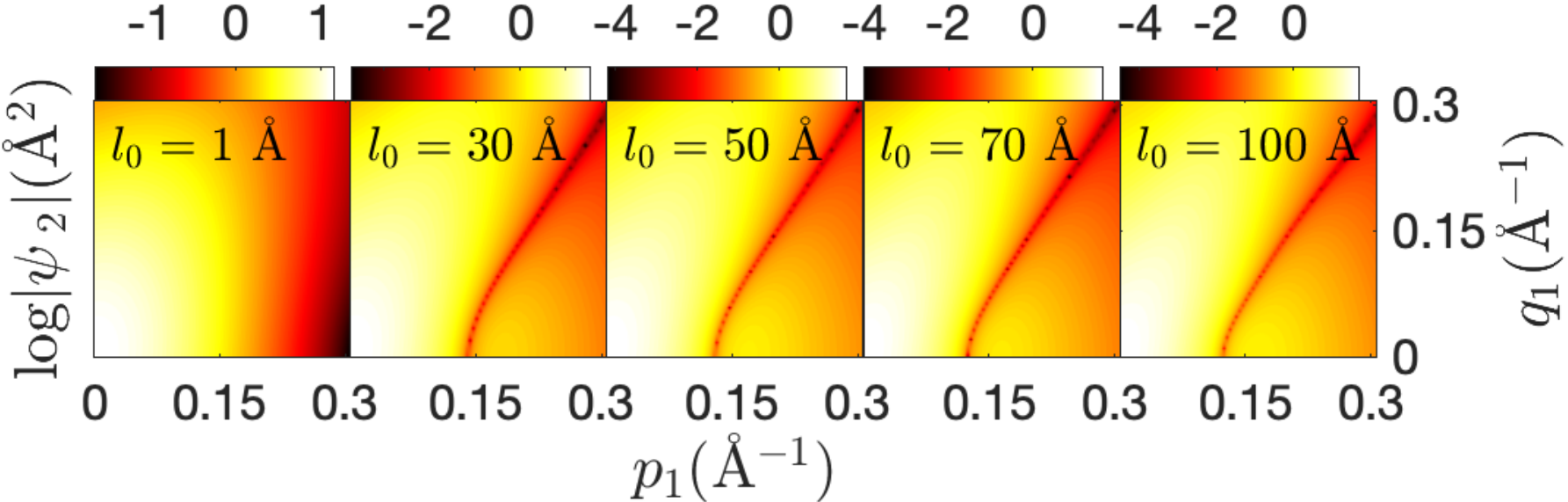}
\caption{The evolution of the Faddeev components (top panels) $\psi_1(p_1,q_1,\phi_1=0)$ and (bottom panels) $\psi_2(p_1,q_1,\phi_1=0)$ obtained for the screening parameter $l_0 = 1, 30, 50, 70, 100$~\AA~in the screening scheme $V_1(q)\to ( 1 - e^{-l_0 q} ) V_{ee}(q)$.}
\label{fig_Faddeev_evolution_1}
\end{figure*}
It is worth noting that in excitonic physics, the electron-hole interaction comprises both attractive screened interaction and repulsive exchange interaction. However, for strongly bound excitons, the exchange interaction generally has a small contribution, as demonstrated in Ref.~\cite{Wang2018}. For the trion, apart from the electron-hole interaction, there is also an exchange term for the electron-electron interaction. However, considering that the electron in the trion is weakly bound to the exciton and associated with small momenta as compared to the reciprocal vector, the exchange terms are presumably much less significant in determining the trion binding energy as compared to the contribution to the exciton energy, which is already small in this case. Despite this, the Faddeev approach to solving the Hamiltonian eigenvalue problem in momentum space is suitable for dealing with non-local exchange terms in a similar way for the exciton Hamiltonian~\cite{Schmidt2003}, which is left for a future study.

\section{Rytova-Keldysh Trion Clusterization} \label{sec:clusterRK}
Trion structure is studied for the screened electron-electron potential $V_1(q)\to ( 1 - e^{-l_0 q} ) V_{ee}(q)$ with $l_0=100$~\AA. We chose this particular model since the electron potential is screened at large distances, which acts together with the natural screening of the exciton interaction with the spectator electron. In this sense, the two effects act coherently, making the trion to be overbinding with an energy of $-207.3$~meV compared to the extrapolated one of $-49.6$~meV. We should keep in mind that features associated with small trion binding energy, with respect to exciton, will be further highlighted towards the converged trion with Rytova-Keldysh potential. Our analysis is based on the screened electron-electron potential, which for the moment, is a limitation of our numerical method applied to the repulsive Rytova-Keldysh potential. Despite that, we study the structure of the trion within the screened model to shed light on its structure and compare it with the 2A+R Yamaguchi model. 
 
We should emphasize that we consider a negatively charged trion with one hole and two electrons, where the two electrons will have the same effective mass. In general, for TMDs, the electrons have the same mass if they belong to the same band/minimum point. This happens for 1) intravalley electrons with the same spin or 2) intervalley and opposite spin electrons, however, they will have the same mass if we neglect the spin-orbit coupling for the conduction band, which is the case in our work.
\begin{figure*}[thb] 
\centering 
\includegraphics[width=.9\linewidth]{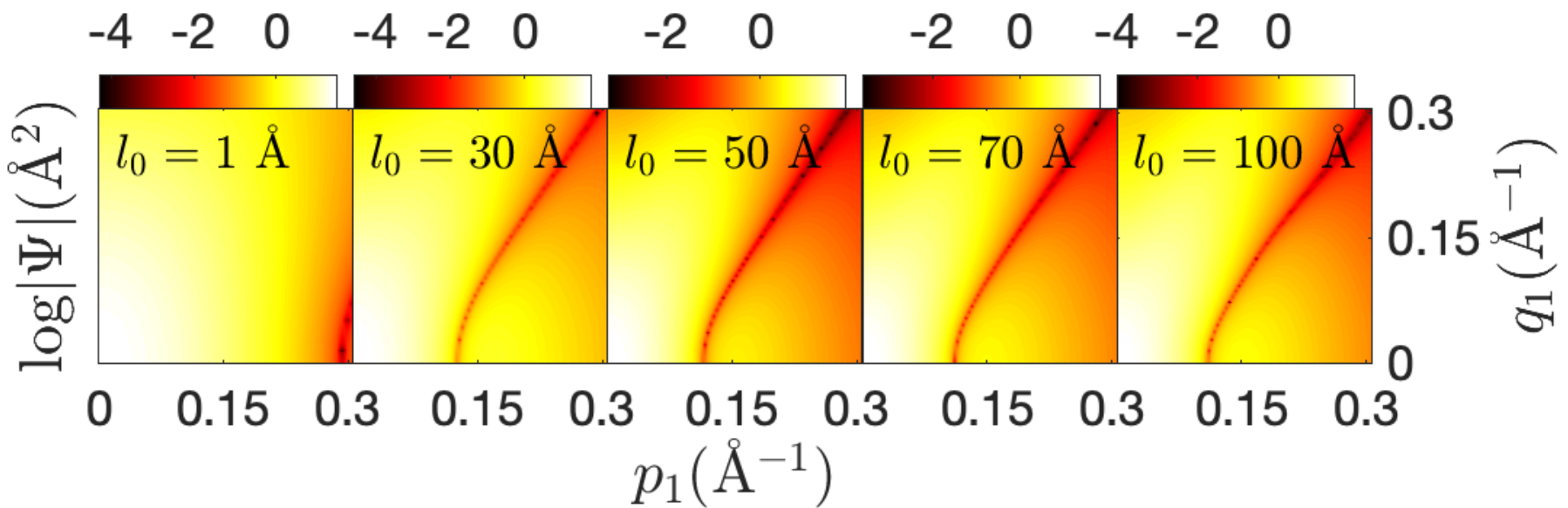}
\\
\includegraphics[width=.9\linewidth]{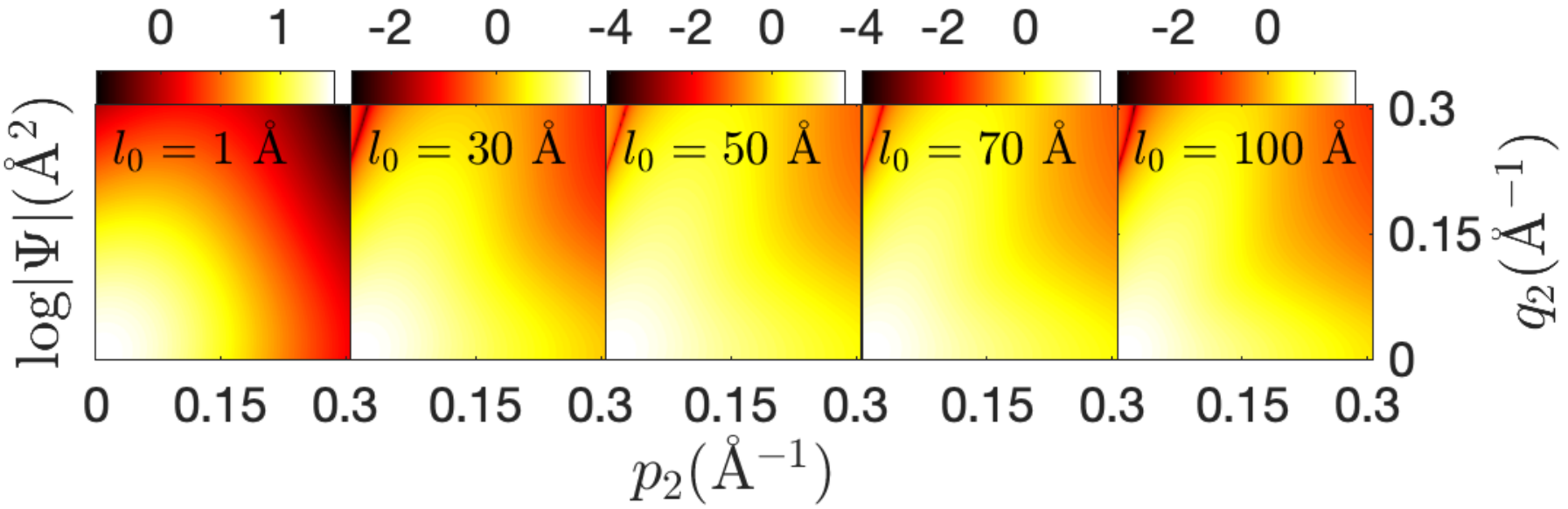}
\caption{The evolution of the total wave function $\Psi(p_1,q_1,\phi_1=0)$ (top panels) and $\Psi(p_2,q_2,\phi_2=0)$ (bottom panels) obtained for the screening parameter $l_0 = 1, 30, 50, 70, 100$~\AA~in the screening scheme $V_1(q)\to ( 1 - e^{-l_0 q} ) V_{ee}(q)$.}
\label{fig_Faddeev_evolution_2}
\end{figure*}
The overlaps between the Faddeev components of the wave function are given in Table~\ref{tab:FaddeevRK}. As expected, the relative normalization of the component $\bera \psi_1 | \psi_1 \ket $, where the hole is the spectator particle of the interacting electron-electron pair, is almost three times smaller than $\bera \psi_2 | \psi_2 \ket=\bera \psi_3 | \psi_3 \ket $. Similarly, we have also observed it in the 2A+R Yamaguchi model (cf. Table~\ref{tab:Yamoverlaps}), which shows a quite small overlap $\bera \psi_1 | \psi_1 \ket $ with respect to the total normalization of the wave function. The configuration where the hole is a spectator of the interacting electron-electron pair is suppressed, favoring the clusterization of the wave function where the electron and hole are close, forming essentially the exciton and a distant spectator electron. We also observe the opposite sign of $\bera \psi_1 | \psi_{2,3} \ket $ with respect to $\bera \psi_2 | \psi_{3} \ket $, as a manifestation of the repulsive interaction between the electrons. The comparison with the 2A+R Yamaguchi potentials overlaps from Table~\ref{tab:Yamoverlaps} with the Rytova-Keldysh results in Table~\ref{tab:FaddeevRK}, showing that the suppression of $\psi_1$ is much more dramatic for the former model. The reason for that is twofold: (i) the relatively smaller difference between the 3B and 2B binding energies from the Yamaguchi model, namely $(E_{3B}-E_{2B})/E_{3B}=0.169$ compared to 0.216 from the Rytova-Keldysh screened model, and (ii) the short-range Yamaguchi potential, while the Rytova-Keldysh potential has a long-range tail. However, the Rytova-Keldysh trion has a considerably smaller 3B binding energy than the resulting one for the screened Rytova-Keldysh electron-electron potential, namely $(E_{3B}-E_{2B})/E_{3B}=0.061$ obtained from the extrapolated value of -802.9~meV in Fig.~\ref{fig_trion_extrapolations}. Therefore, we expect a more evident clusterization of the exciton within the trion. 
\begin{figure*}[thb] 
\centering
\includegraphics[width=.9\linewidth,angle=0]{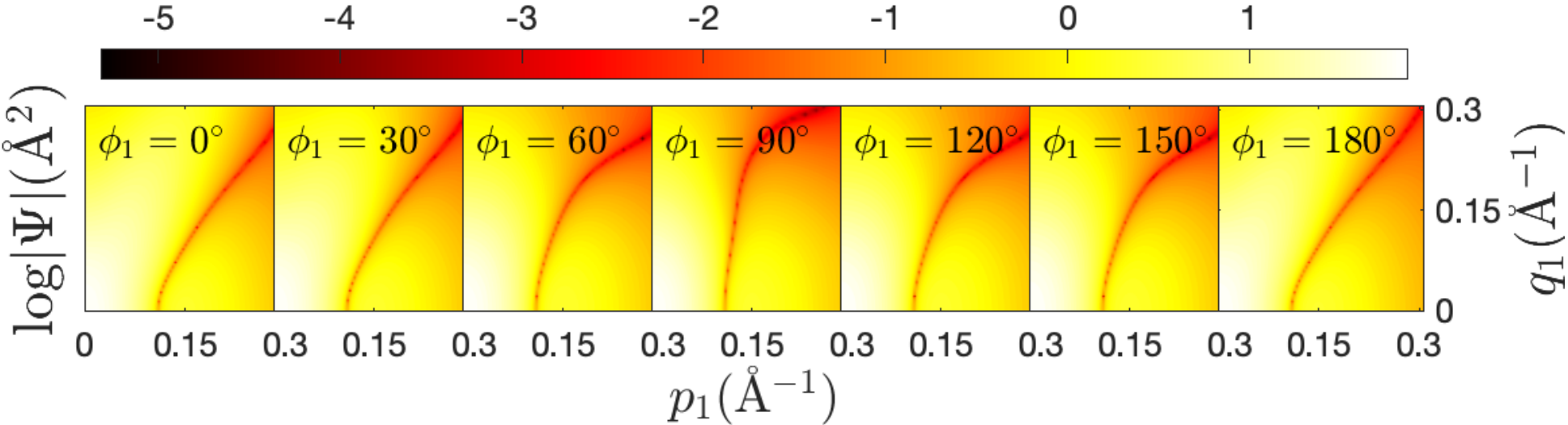}
\caption{The evolution of the total wave function $\Psi(p_1,q_1,\phi_1)$ with respect to the angle $\phi_1$ for a fixed screening parameter $l_0 =100$~\AA~in the screening scheme $V_1(q)\to ( 1 - e^{-l_0 q} ) V_{ee}(q)$.}
\label{fig:WFangles}
\end{figure*}
\begin{figure}[thb] 
\centering
\begin{tabular}{cc}
\includegraphics[width=4.1cm,angle=0]{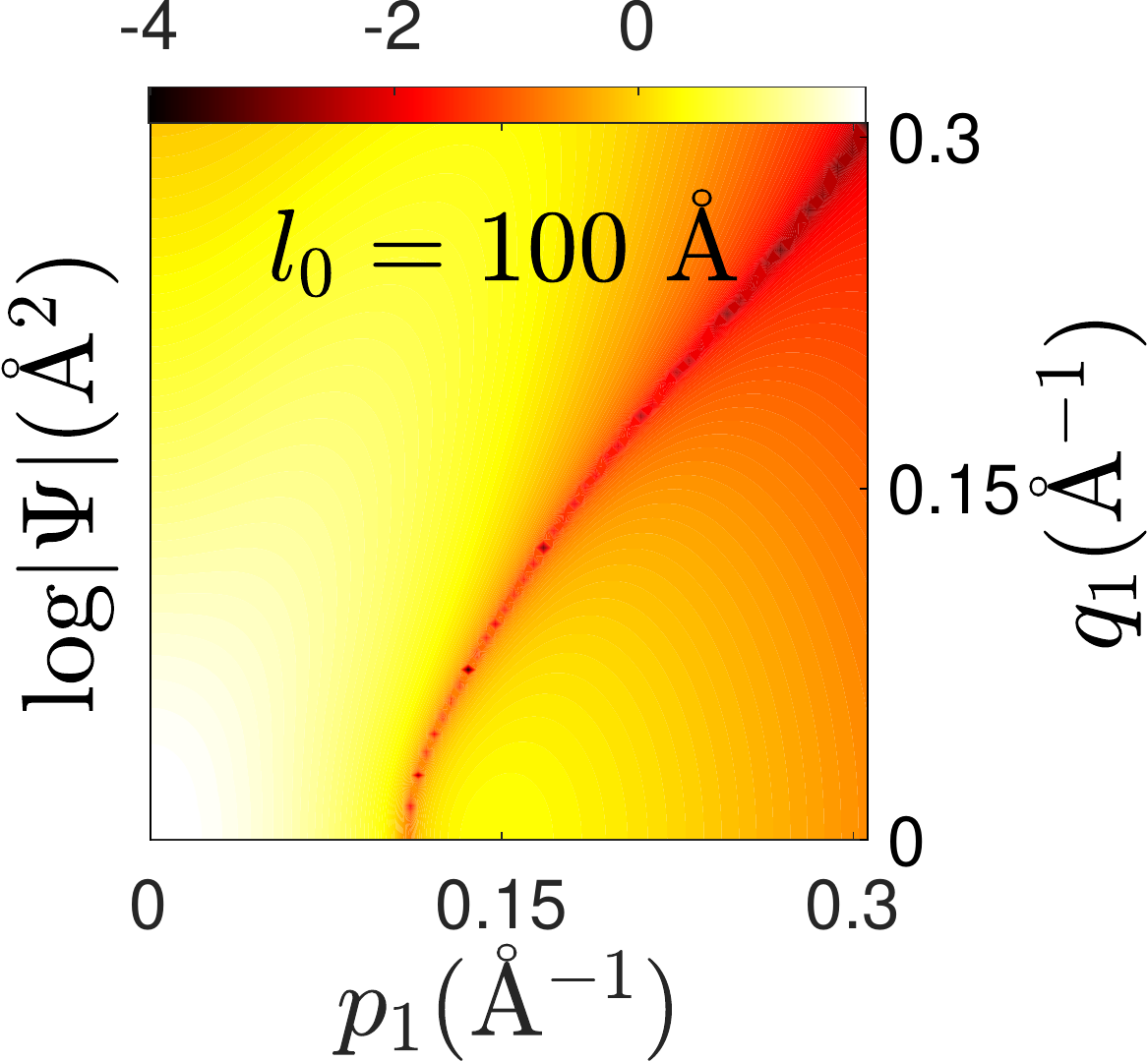}
&
\includegraphics[width=4.1cm,angle=0]{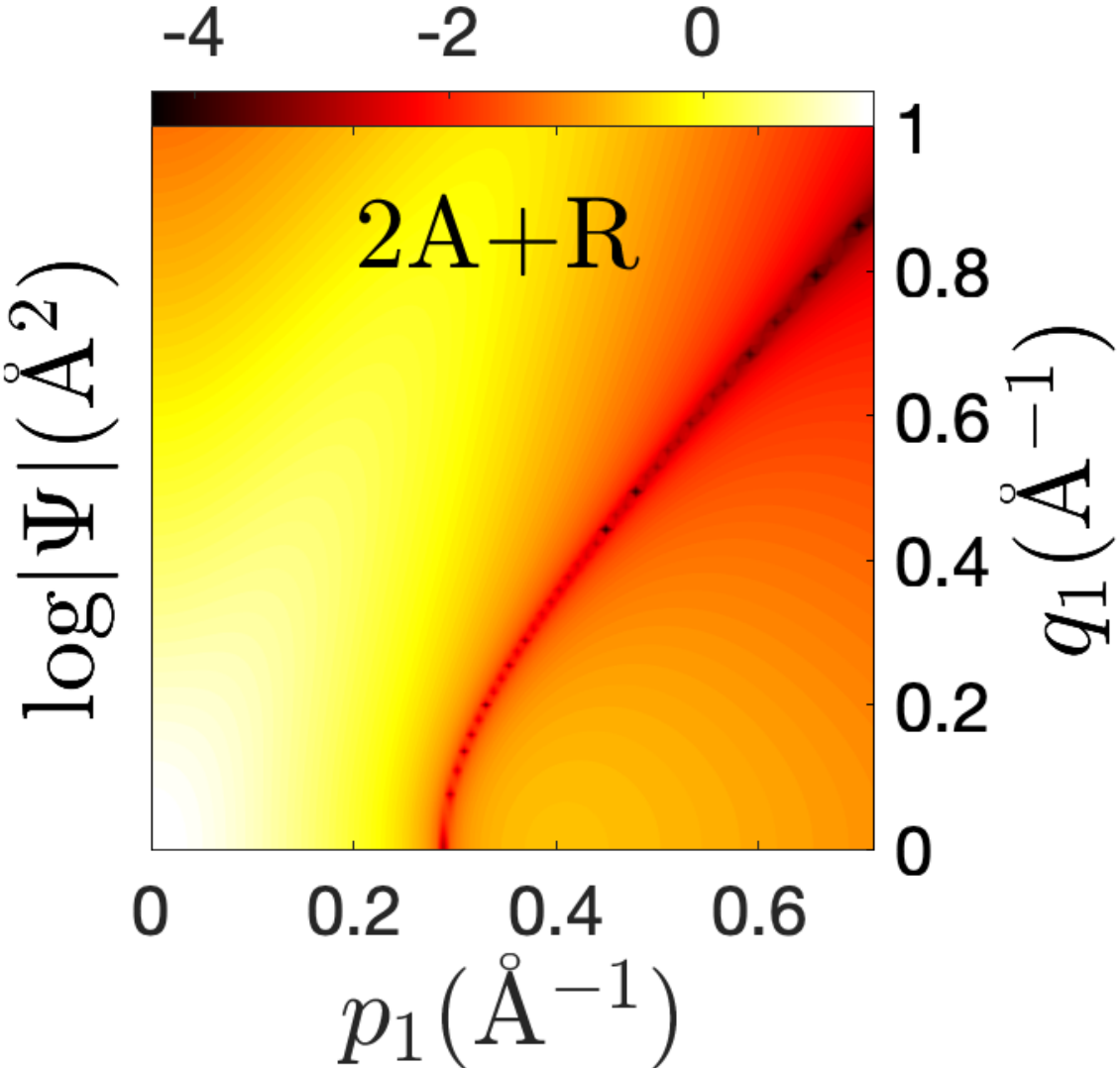}
\end{tabular}
\caption{Comparison between trion wave function calculated with Rytova-Keldysh potential (left panel) and re-scaled Yamaguchi potential model (right panel) for $\phi_1=0$. } \label{fig:WFYAMRK}
\end{figure}
In Fig.~\ref{fig_Faddeev_evolution_1}, it is shown the results for the Faddeev components $\psi_1(p_1,q_1,\phi_1=0)$ (top panel) and $\psi_2(p_2,q_2,\phi_2=0)=\psi_3(p_3,q_3,\phi_3=0)$ (bottom panel), where the momenta are defined in terms of $p_1$ and $q_1$ according to Eq.~\eqref{eq:q2p2q3p3}. The calculations were done for $l_0 = 1, 30, 50, 70, 100$~\AA~(from left to right panels) with the electron-electron screened potential $V_1(q)\to ( 1 - e^{-l_0 q} ) V_{ee}(q)$. As $l_0$ increases, the trion binding decreases and the electrons are pushed away from each other. At the same time, one of them has a hole in the vicinity region of the exciton size as expressed by the dominant configurations, namely $\psi_2\sim$~[3(e)1(h)]---2(e) and $\psi_3\sim$~[1(h)2(e)]---3(e) (the square brackets represent the exciton), while $\psi_1(p_1,q_1,0)$ just reflects the short-range repulsion, spreading $q_1$ and $p_1$ to larger values, as depicted in the top panels of Fig.~\ref{fig_Faddeev_evolution_1}. For small values of the screening parameter $l_0$, in which the long-range tail of the repulsive potential is damped, we observe, analyzing the momentum distribution of $\psi_1(p_1,q_1,0)$, larger trion bindings tending to be more symmetrical in $p_1$ and $q_1$. As a consequence, it implies a geometric configuration where the hole is equally separated from the two electrons, disfavoring the cluster structure against a more symmetrical configuration, schematically like e--h--e. The node appears in $\psi_1$ at larger values of $p_1$ for large $l_0$ values, as the repulsion is increased and it is intense at the short range. 

In the bottom panel of Fig.~\ref{fig_Faddeev_evolution_1}, the evolution of $\psi_2(p_2(\bp_1,\bq_1),q_2(\bp_1,\bq_1),\phi_2=0)$ for $\phi_1=0$ with $l_0$ is shown in the ($p_1\times q_1$)--plane. As already observed in the top panel of Fig.~\ref{fig_Faddeev_evolution_1} for $\psi_1$, as $l_0$ increases, the configuration migrates from the symmetrical situation, e--h--e, to a cluster one, [3(e)1(h)]---2(e). The node line becomes evident together with the cluster structure when the long-range screening is reduced, and the electron becomes weakly bounded with respect to the exciton. Higher amplitude values of the wave function are found for $p_1\propto q_1$, when the cluster structure dominates, as already discussed for the 2A+R Yamaguchi 3B model.

The total wave function is presented in Fig.~\ref{fig_Faddeev_evolution_2} in the $p_1\times q_1$ (top panel) and $p_2\times q_2$ (bottom panel) planes for several values of $l_0$ from 1 to 100~\AA~with $\phi_1=\phi_2=0$. The reduction of the screening at large distances turns the pattern similar to the one observed for $\psi_2$ in the ($p_1\times q_1$)--plane [cf. bottom panels in Fig.~\ref{fig_Faddeev_evolution_1}], more evident due to its dominance over $\psi_1$, reminding that $\psi_2\equiv\psi_3$ from the symmetric configuration of the two electrons, which have to be in a singlet spin state or an antisymmetric combination of different valley states. In the bottom panels of Fig.~\ref{fig_Faddeev_evolution_2}, results for the total wave function are presented in the ($p_2\times q_2$)--plane, where again, the more symmetric e--h--e configuration dominates at the strong trion binding and weaker repulsion between the electrons. By reducing the screening of the repulsive potential, the electron turns to be weakly bound to the exciton, and the system presents an evident cluster structure with the coherent superposition of the two configurations [3(e)1(h)]---2(e) and [1(h)2(e)]---3(e). As a function of $p_2$ and $q_2$, the total wave functions, demonstrated in the top panels of Fig.~\ref{fig_Faddeev_evolution_2}, exhibit two branches of higher probability density, namely for $p_2\propto q_2$ and for $q_2$ small with $p_2$ spreading in the region shown in the figure. This behavior was found in the 2A+R Yamaguchi model and is associated with the cluster structure, already discussed in detail in Sec.~\ref{sec:clusteryamaguchi}.

In Fig.~\ref{fig:WFangles}, the angular dependence in $\phi_1$ is explored for $l_0=100$~\AA. Note that, at $\phi_1=\theta$ or $\phi_1=180^o-\theta$, the results are the same due to the symmetry of the wave function by exchanging the momentum of the electrons. The configuration space wave function is symmetrical by the exchange of the electrons once the antisymmetry is ensured by the spin state. The slope of the node line is deformed when $\phi_1$ changes between $\bp_1$ and $\bq_1$, and becomes more elongated at 90$^\circ$. The node line format is basically kept regardless of the $\phi_1$-parameters, which in turn reflects the situation where the e-h-e system has a hole with small momentum with respect to the center of mass of the electron pair. It can be clearly seen that the zero of the wave function in the ($p_1\times q_1$)--plane starts at $q_1=0$.

Finally, in Fig.~\ref{fig:WFYAMRK}, we directly compare the trion wave function computed with both Rytova-Keldysh (left panel) and 2A+R Yamaguchi (right panel) models. For that, we performed a re-scaling of the Yamaguchi separable potential model to physical units of the exciton and trion. The results for the exciton and trion were obtained for the Yamaguchi model in units of $\hbar=m=1$. Turning to physical units we have that $$E_{2B}=-\frac{\hbar^2}{m_e}\lambda^2\, 0.1 \quad \text{and}\quad E_{3B}=-\frac{\hbar^2}{m_e}\lambda^2\,0.12034\, ,$$ 
and to obtain the dimensional constant $\lambda$ we use the exciton binding energy of $E_{2B}=-753$~meV, and $m=(m_e+m_h)/2=0.505\,m_0$, which gives $\lambda=0.706576$~
\AA$^{-1}$. The trion binding energy in the 2A+R Yamaguchi model in meV units is $E_t=153.225$~meV, comparable with 207.26~meV from the regulated repulsive Rytova-Keldysh potential with $l_0 =100$~\AA{} given in Table~\ref{table_RK_att_rep_expectation}.
After the re-scaling to physical units, the comparison shows essentially the same structure of the trion wave function resulting from calculations with the Rytova-Keldysh and 2A+R Yamaguchi potentials. Remarkably the node line has the same form, and the region where the wave function attains the highest values is to the left of the node line with $p_1\propto q_1$. The node line is shifted to larger values of $p_1$ for the 2A+R Yamaguchi model with respect to the Rytova-Keldysh potential, which should reflect the short-range nature of the Yamaguchi potential contrasting with the Rytova-Keldysh potential.

\section{Summary}
\label{sec:summary} 
Our work can be helpful for theoretical-computational research on trions on the following main aspects: (i) we provided a general formulation of the Faddeev equations to compute the wave function for a 2D trion in momentum space; (ii) we developed a method for the accurate calculation of the trion binding energy in freestanding monolayer MoS$_2$ with different regularization schemes for the repulsive electron-electron Rytova-Keldysh potential, with the numerical precision checked by computing the expectation value of the Hamiltonian; and (iii) we analyzed in detail the degree of clusterization of the weakly bound trion state through the momentum distributions of the total wave function and its Faddeev components. 

The repulsive electron-electron Rytova-Keldysh potential posed a numerical challenge in solving the Faddeev equations. To overcome that, we have devised two different regularization functions applied only to the repulsive term, which turns the trion weakly bound with respect to the exciton. The two different choices for the regularization functions 
were chosen as (i) $e^{-l_0q}$ ($7<l_0<25$~\AA), that acts in the high momentum transfer region, and (ii) $1-e^{-l_0q}$ ($1<l_0<100$~\AA), that acts in the low momentum transfer region. The results were then extrapolated to $l_0^{-1}\to 0$ in the former case and to $l_0\to 0$ in the latter one. The extrapolation results were good in one part in $10^4$, resulting in a prediction of the trion energy of $-49.3(1)$~meV for monolayer MoS$_2$. Our result lies in the range of the experimental results for suspended samples 44-80~meV as reported in Refs.~\cite{lin2014dielectric, Lin2019}. It can not be compared with the results where the MoS$_2$ is deposited on a substrate, where the trion binding energy lies in the range of $20-43$~meV~\cite{Mak_2012, Ross2013, Soklaski2014, Zhang2015}, since the interaction becomes weaker due to screening, affecting both the exciton and trion complexes. The value of $-49.3(1)$~meV is within the previous numerical results and particularly close to the converged negatively charged intralayer trion binding energy computed within an ab-initio many-body theory, which was found to be $58$~meV~\cite{Druppel_2017} with the exciton binding energy of $-0.76$~eV.

Furthermore, we have analyzed the structure of the trion wave function by decomposing it into its Faddeev components for both the Rytova-Keldysh and Yamaguchi potentials. Despite the fact that the two interactions have different large distance tails, we have observed qualitative similarities in the wave functions. Both trion models showed a remarkable  dominance of the [eh]--e configurations, corresponding to the Faddeev components of the wave function where the electron is a spectator. We should observe that the electrons are in a symmetric configuration by exchanging their spatial coordinates and are considered to be in a spin singlet state or an antisymmetric combination of different valley states.

We expect that without regulating the repulsive Rytova-Keldysh potential, the dominance of the strong cluster structure [eh]--e would be more evident, as the trion will be even more weakly bound, as indicated by the extrapolated results. Therefore, our study suggests that a realistic Rytova-Keldysh potential calculation, with the small relative exciton and trion splitting, can profit from the cluster structure and use it to build more accurate methods,  eventually relying on the dominant exciton structure together with a simplified potential that contains the low-energy electron-electron continuum information.

\begin{acknowledgments}
We thank Charlotte Elster for the helpful discussion on the Coulomb interaction screening. 
This work is a part of the project INCT-FNA proc. No. 464898/2014-5.
The work of M.~R.~H. was supported by the National Science Foundation under Grant No. NSF-PHY-2000029 with Central State University.
M.~R.~H. also thanks the Ohio Supercomputer Center (OSC) for the use of their facilities under Grant No. POS0104.
K.~M. acknowledges a Ph.D. scholarship from the Brazilian agency CNPq (Conselho Nacional de Desenvolvimento Cient\'ifico e Tecnol\'ogico). 
K.~M., T.~F., A.~J.~C., and D.~R.~C. were supported by CNPq Grant No.~400789/2019-0, 308486/2015-3, 315408/2021-9, and 313211/2021-3, respectively. 
A.~J.~C. and T.~F. acknowledge Funda\c{c}\~ao de Amparo \`a Pesquisa do Estado de S\~ao Paulo (FAPESP) under Grant No.~2022/08086-0 and Thematic Projects 2017/05660-0 and 2019/07767-1, respectively.

\end{acknowledgments}

\appendix

\section{Momentum space representation of Faddeev equations}\label{App_momentum_Faddeev}

\begin{figure}[H]
\centering
\begin{tikzpicture}[line cap=round,line join=round,>=triangle 45,x=0.9cm,y=0.9cm]
\coordinate (O) at (0,0);
\coordinate (A) at (7,0);
\coordinate (B) at (7,0);
\draw [->,line width=1pt] (O) -- (A)node [right]{$x$};
\draw [->,line width=1pt,dashed] (O) -- (0:6)coordinate(C) node [below]{$\bp_i$};
\draw [->,line width=1pt] (O) -- (90:6)node [right]{$y$};
\draw [->,line width=1pt,dashed] (O) -- (40:6)coordinate(C) node [right]{$\bp'_i$};
\draw [->,line width=1pt,dashed] (O) -- (60:6)coordinate(B) node [right]{$\bq_i$};
\pic [draw, ->, "$\phi'_{i}$", angle eccentricity=1.1,angle radius=3cm] {angle = A--O--C};
\pic [draw, ->, "$\phi_i$", angle eccentricity=1.12,angle radius=2cm] {angle = A--O--B};
\pic [draw, ->, "$\phi_{q_i,p'_i}$", angle eccentricity=1.1,angle radius=3.5cm] {angle = C--O--B};
\end{tikzpicture}
\caption{The coordinate system for the solution of the coupled Faddeev integral equations \eqref{Faddeev_1.2}.}
\label{fig_coordinate_1}
\end{figure}
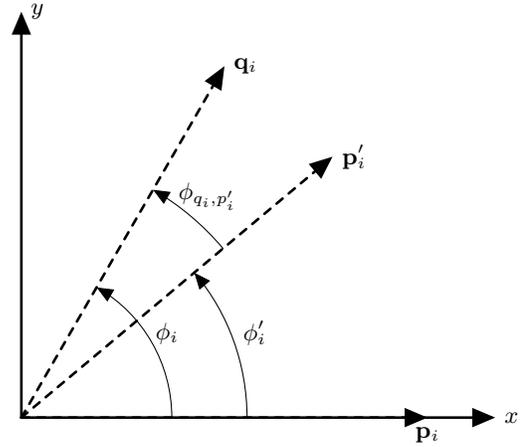

The Faddeev components, i.e., the projections of the coupled Faddeev equations \eqref{coupled_Faddeev_operator} on the 3B basis states $\vert \bp_i \bq_i \ket $, can be written as
\begin{equation}
\label{psi_completness}
 \bera \bp_i \bq_i \vert \psi_i \ket =G_{0}^{(i,jk)}(E,p_i,q_i)
 \bera \bp_i \bq_i \vert t_i \Big[\vert\psi_j \ket +
 \vert\psi_k \ket 
 \Big]\,. 
\end{equation}
By inserting the completeness relation of Eq.~\eqref{completeness_relation} into Eq.~\eqref{psi_completness}, it leads to 
\begin{eqnarray}
 \bera \bp_i \bq_i \vert \psi_i \ket &=&
G_{0}^{(i,jk)}(E,p_i,q_i)
\int d^2p'_i \int d^2q'_i 
\cr
&\times&
 \bera \bp_i \bq_i \vert t_i \vert \bp'_i \bq'_i \ket \bigl[ \bera \bp'_i \bq'_i\vert \psi_j \ket +
 \bera \bp'_i \bq'_i\vert\psi_k \ket 
 \bigr] , \quad
 \label{Faddeev_1}   
\end{eqnarray}
where 
\begin{eqnarray}
G_0^{(i,jk)}(E,p_i,q_i)= \dfrac{1}{E-\dfrac{p_i^2}{2\mu_{jk}} - \dfrac{q_i^2}{2\mu_{i,jk}}},
\label{G.1}
\end{eqnarray}
and
\begin{eqnarray}
 \bera \bp_i \bq_i \vert t_i \vert \bp_i' \bq_i' \ket &=& \delta(\bq_i-\bq'_i)
 \bera \bp_i\vert t_i \vert \bp_i' \ket . 
 \label{1_t1}
\end{eqnarray}
In order to evaluate Eq.~\eqref{Faddeev_1}, we need to compute $ \bera \bp'_i \bq'_i\vert\psi_j \ket $ and 
 $ \bera \bp'_i \bq'_i\vert\psi_k \ket $, where by inserting a completeness relation, one obtains
\begin{eqnarray}
 \bera \bp'_i \bq'_i\vert\psi_j \ket &=&
\int d^2p''_j \int d^2q''_j \bera \bp_i' \bq_i' 
\vert \bp''_j \bq''_j \ket \bera \bp''_j \bq''_j \vert
\psi_j \ket 
\cr 
&=& 
 \bera \bp_j(\bp'_i,\bq'_i), \bq_j(\bp'_i,\bq'_i) \vert
\psi_j \ket ,
\cr
 \bera \bp'_i \bq'_i\vert\psi_k \ket &=&
\int d^2p''_k \int d^2q''_k \bera \bp_i' \bq_i' 
\vert \bp''_k \bq''_k \ket \bera \bp''_k \bq''_k \vert
\psi_k \ket 
\cr
&=& 
 \bera \bp_k(\bp'_i,\bq'_i), \bq_k(\bp'_i,\bq'_i) \vert
\psi_k \ket ,
\label{eq_psi_jk}
\end{eqnarray}
where the relation between different Jacobi momenta is given by
\begin{eqnarray}\label{eq:A6}
\bp_i(\bp_j,\bq_j)&\equiv& \pmb{\mathcal{P}}_{ij} (\bp_j,\bq_j) =\alpha_{ij} \bp_j +\beta_{ij} \bq_j ,
\\
\bq_i(\bp_j, \bq_j) &\equiv&\pmb{\mathcal{Q}}_{ij} (\bp_j, \bq_j) = \gamma_{ij} \bp_j+\eta_{ij}\bq_j,\label{eq:A7}
\end{eqnarray}
with 
\begin{eqnarray}
&& \alpha_{ij} = -\dfrac{m_j}{m_{jk}}, \,
\beta_{ij} = \mathcal{E}_{ij} \dfrac{m_k \, m_{ijk}}{ m_{ik} m_{jk} },
\cr 
&& \gamma_{ij} = -\mathcal{E}_{ij},\,
\eta_{ij} = -\dfrac{m_i}{m_{ik}},
\cr
&& m_{ij} = m_i+m_j,\quad
m_{ijk}= m_i +m_j+m_k, 
\cr 
&& \mathcal{E}_{ij}= 
 \begin{cases}
  1 & \text{for cyclic permutation}\\
  -1 & \text{for anti-cyclic permutation} 
 \end{cases}. 
 \label{eq:A8}
\end{eqnarray}
By using Eqs.~\eqref{1_t1} and \eqref{eq_psi_jk} we can rewrite Eq.~\eqref{Faddeev_1} as
\begin{eqnarray}
\psi_i (\bp_i,\bq_i ) &=&
G_{0}^{(i,jk)}(E,p_i,q_i)\int d^2p'_i 
\, t_i(\bp_i,\bp_i') 
\cr
 &\times& 
\Biggl[
 \psi_j \biggl ( \pmb{\mathcal{P}}_{ji}(\bp'_i,\bq_i), \pmb{\mathcal{Q}}_{ji}(\bp'_i,\bq_i) \biggr) \cr
&& +
 \psi_k \biggl ( \pmb{\mathcal{P}}_{ki}(\bp'_i,\bq_i), \pmb{\mathcal{Q}}_{ki}(\bp'_i,\bq_i) \biggr)
 \Biggr].
 \label{Faddeev_1.2}
\end{eqnarray}
To solve the coupled 2D Faddeev integral equations, {\it i.e.} Eq.~\eqref{Faddeev_1.2}, as shown in Fig.~\ref{fig_coordinate_1}, we choose a coordinate system where $\bp_i$ is parallel to the $x-$axis, $\bp'_i$ and $\bq_i$ are free in the 2D space.
So the angle variables will be as
\begin{eqnarray}
(\hat{\bq_i} , \hat{\bp_i}) &=& \phi_i, \\
(\hat{\bp'_i} , \hat{\bp_i}) &=& \phi'_i,\\
(\hat{\bq_i} , \hat{\bp'_i}) &=& \phi_{q_i,p'_i}=\phi_i- \phi'_i .
\end{eqnarray}
The shifted momentum and angle variables are defined as
\begin{eqnarray}
\mathcal{P}_{ij}&\equiv&\mathcal{P}_{ij}(\bp'_j,\bq_j) =\vert \alpha_{ij} \bp'_j +\beta_{ij} \bq_j \vert
= \sqrt{(\mathcal{P}_{ij}^{X})^2+( \mathcal{P}_{ij}^{Y})^2} , \cr
 \mathcal{Q}_{ij}&\equiv&\mathcal{Q}_{ij}(\bp'_j,\bq_j) = \vert \gamma_{ij} \bp'_j+\eta_{ij}\bq_j \vert= \sqrt{(\mathcal{Q}_{ij}^{X})^2+( \mathcal{Q}_{ij}^{Y})^2}, \cr
\phi_{ij} &=& (\hat{\mathcal{P}_{ij}},\hat{\mathcal{Q}_{ij}}) = {\text{atan}2}(det,dot), \quad
 0 < \phi_{ij} < 2\pi,\label{eq:shifted}
\end{eqnarray}
where
\begin{eqnarray}
\mathcal{P}_{ij}^{X} &=& \alpha_{ij} p'_{j} \cos(\phi'_j)+\beta_{ij}q_j \cos(\phi_j), \cr
\mathcal{P}_{ij}^{Y} &=& \alpha_{ij}p'_{j}\sin(\phi'_j) + \beta_{ij}q_j \sin(\phi_j), \cr
\mathcal{Q}_{ij}^{X} &=& \gamma_{ij} p'_j \cos(\phi'_j) +\eta_{ij}q_j \cos(\phi_j), \cr
 \mathcal{Q}_{ij}^{Y} &=& \gamma_{ij}p'_j\sin(\phi'_j) + \eta_{ij}q_j \sin(\phi_j),
\end{eqnarray}
and
\begin{eqnarray}
 \begin{cases}
  det = \mathcal{P}_{ij}^{X} \cdot \mathcal{Q}_{ij}^{Y} - \mathcal{P}_{ij}^{Y} \cdot \mathcal{Q}_{ij}^{X},\\
  dot = \mathcal{P}_{ij}^{X} \cdot \mathcal{Q}_{ij}^{X} + \mathcal{P}_{ij}^{Y} \cdot \mathcal{Q}_{ij}^{Y}.
 \end{cases} 
\end{eqnarray}
%

\section{3B wave functions in momentum space}
\label{App_3B_WF_momentum}
3B wave function is given as
\begin{eqnarray}
\Psi = \sum_{i=1}^{3} \psi_i,
\end{eqnarray}
where 
\begin{equation}
\bera \bp_i \bq_i \vert \Psi \ket= \bera \bp_i \bq_i \vert \psi_i \ket +
\bera \bp_i \bq_i \vert \psi_j \ket
+
\bera \bp_i \bq_i \vert \psi_k \ket.
\end{equation}
Using Eq.~\eqref{eq_psi_jk} and the coordinate system defined in Fig.~\ref{fig_coordinate_1}, one has that
\begin{eqnarray}
&& \Psi(p_i,q_i,\phi_i) = \psi_i(p_i,q_i,\phi_i)
\cr && +
\psi_j \biggl ( \mathcal{P}_{ji}(p_i,q_i,\phi_i), \mathcal{Q}_{ji}(p_i,q_i,\phi_i),\phi_{ji}(p_i,q_i,\phi_i) \biggr)
\cr
&& +
\psi_k \biggl ( \mathcal{P}_{ki}(p_i,q_i,\phi_i), \mathcal{Q}_{ki}(p_i,q_i,\phi_i),\phi_{ki}(p_i,q_i,\phi_i) \biggr). \quad
\end{eqnarray}
The 3B wave function is normalized as
\begin{eqnarray}
\bera \Psi \vert \Psi\ket &=& \sum_{i=1}^3 \bera \Psi \vert \psi_i \ket 
\cr
 &=& 
2 \pi \int_0^{\infty} dp_i \, p_i 
\int_0^{\infty} dq_i \, q_i \int_{0}^{2\pi} d\phi_i \,
\Psi^2(p_i,q_i,\phi_i) 
\cr
 &=& 
2 \pi \int_0^{\infty} dp_i \, p_i 
\int_0^{\infty} dq_i \, q_i \int_{0}^{2\pi} d\phi_i \,
\Psi(p_i,q_i,\phi_i) 
\cr
&\times&
\biggl[ \psi_i(p_i,q_i,\phi_i)
\cr
&& +
\psi_j( \mathcal{P}_{ji}(\bp_i,\bq_i), \mathcal{Q}_{ji}(\bp_i,\bq_i),\phi_{ji})
\cr
&& +
\psi_k( \mathcal{P}_{ki}(\bp_i,\bq_i), \mathcal{Q}_{ki}(\bp_i,\bq_i),\phi_{ki})
\biggr]
=1.
\label{eq.normalization_3B}
\end{eqnarray}

\begin{table*}[thb]
\centering
\caption{The convergence of trion ground state binding energies (in meV) as a function of the number of mesh points for Jacobi momenta $N_p = N_q$ obtained for different values of screening parameter $l_0$ for screening electron-electron interactions $V(q) \to e^{-l_0\,q} V_{ee}(q)$ (upper panel) and $V(q) \to ( 1 - e^{-l_0 q} ) V_{ee}(q)$ (lower panel). The last row of each panel shows the extrapolation of trion energy eigenvalues to an infinite number of mesh points.
}
\begin{tabular}{c|c|c|c|c|c|c|c|c|c|c}
\hline
&\multicolumn{10}{c}{$l_0(\AA) $}
\\ \cline{2-11}
 $N_p = N_q$ &$25$& $20$ & $17$& $15$ & $13$ & $11$ & $10$ &$9$ &$8$ &$7$\\
   \hline
 $200$ & $-1275.6$ & $-1231.1$ & $-1197.7$ & $-1171.8$ & $-1142.5$ &$ -1109.0$ &$-1090.5$ & $-1070.8$ & $-1049.8$ & $-1027.6$\\
   \hline
 $250$ & $ -1258.7$ & $-1214.2$ & $-1180.8$ & $ -1155.0$ & $-1125.6$ &$-1092.2$ &$-1073.7$ & $-1054.1$ & $-1033.0$ & $-1011.5$\\
   \hline
 $300$ &$-1248.3$ & $-1203.8$ & $-1170.4$ & $-1144.6$ & $-1115.3$ &$-1081.8$ &$-1063.4$& $-1043.7$ & $-1022.8$ & $-1000.7$\\
   \hline
$350$ & $-1241.3$ & $-1196.8$ & $-1163.5$ & $-1137.7$ & $-1108.3$ &$-1074.9$ &$-1056.5$& $-1036.8$ & $-1016.0$ & $-993.7$ \\
   \hline
$ N_p, N_q \to \infty$& $-1195.1$ & $-1150.6$ & $-1117.3$ & $-1091.5$ & $-1062.2$ &$-1028.9$ &$-1010.6$& $-991.0$ & $-970.3$ & $-948.0$ 
\\
\hline
&\multicolumn{10}{c}{$l_0 (\AA) $}
\\ \cline{2-11}
 $N_p = N_q$ &$1$ &$5$ &$10$ &$15$ &$20$ & $30$ & $50$& $70$ &$90$ & $100$\\
 \hline
 $200$ &$-1588.7$ &$-1447.8$&$-1346.2$ &$-1279.4$ & $-1231.0$ & $-1165.0$ & $-1090.8$ & $-1049.3$ & $-1023.3$ & $-1014.8$ \\
   \hline
 $250$ &$-1558.6$ &$-1418.8$& $-1318.1$ &$-1251.8$ & $-1203.8$ & $-1138.2$ & $-1064.7$ & $-1023.6$ & $-998.3$ & $-988.8$ \\
   \hline
 $300$ &$-1540.0$ &$-1400.9$&$-1300.7$ &$-1234.8$ & $-1187.0$ & $-1121.7$& $-1048.4$ & $-1007.1$ & $-981.1$ & $-972.3$ \\
   \hline
 $350$&$-1527.5$ &$-1388.9$ & $-1289.1$ &$-1223.3$ & $-1175.7$ & $-1110.6$& $-1037.5$ & $-996.3$ & $-970.0$ & $-960.6$ \\
   \hline
$N_p, N_q \to \infty$&$-1444.8$ &$-1309.6$& $-1212.0$ &$-1147.7$ & $-1101.2$ & $-1037.3$& $-965.6$ & $-924.8$ & $-898.2$ & $-888.0$ \\
\hline
\end{tabular}
\label{table_trion_l0}
\end{table*}
%

\section{Expectation value of 3B Hamiltonian}
\label{App_expectations}
By having the 3B wave function, the expectation value of 3B Hamiltonian $ \bera H \ket $ can be obtained as
\begin{equation}
 \bera \Psi \vert H \vert \Psi \ket = \bera \Psi \vert H_0 \vert \Psi \ket + \bera \Psi \vert V \vert \Psi \ket ,
 \label{eq:expectation}
\end{equation}
where
\begin{eqnarray}
&& \bera \Psi \vert H_0 \vert \Psi \ket 
= 
 2 \pi \int_0^{\infty}
dp_i \, p_i \int_0^{\infty} dq_i \, q_i 
\left ( \dfrac{p_i^2}{2\mu_{jk}} + \dfrac{q_i^2}{2\mu_{i,jk}} \right) 
 \cr && \times \int_{0}^{2\pi} d\phi_i \, \Psi^2(p_i,q_i,\phi_i) .
\label{eq:expectation_H0}
\end{eqnarray}
and
\begin{eqnarray}
&& \bera \Psi \vert V \vert \Psi\ket = \sum_{i=1}^3\bera \Psi \vert V_i \vert \Psi \ket 
\cr && =
 2\pi \sum_{i=1}^3 \int_0^{\infty} dp_i p_i \,\int_0^{\infty} dq_i q_i \, \int_0^{2\pi} d\phi_i \, \int_0^{\infty} dp'_i \, p'_i \, \int_0^{2\pi} d\phi'_i
 \cr && \times
\Psi(p_i,q_i,\phi_i) \,
V_i(p_i,p'_i,\phi'_i) \,
\Psi(p'_i,q_i,\phi_i- \phi'_i).
\end{eqnarray}
The matrix elements of non-PW potentials can be obtained from the summation of PW components as
\begin{equation}
V(p_i,p'_i, \phi'_i )= \frac{1}{2\pi} \,\sum^{\infty}_{m=0} \varepsilon_m cos(m\phi'_i)\, V_m(p_i,p'_i).
\label{Eq.V_pw_nonpw} 
\end{equation} 
%

\section{Verification of the 3B Schr\"odinger equation in 2D}
\label{App_test_Schrodinger}
The Schr\"odinger equation for the bound state of three particles is given by 
\begin{equation}
E_t \vert \Psi \ket = H\vert \Psi \ket = (H_0 +V_i +V_j+ V_k)\vert \Psi \ket .
\end{equation}
Using the three different sets of Jacobi momenta in momentum space, we obtain
\begin{eqnarray}
&& E_t\Psi(\bp_i,\bq_i) = \left[ \dfrac{p_i^2}{2\mu_{jk}} + \dfrac{q_i^2}{2\mu_{i,jk}}\right] \,\Psi(\bp_i,\bq_i)
\cr
&& + \int d^2p'_i \,V_i(\bp_i,\bp'_i)\,\Psi(\bp'_i,\bq_i)
\cr
&& + \int d^2p'_j\, V_j (\pmb{\mathcal{P}}_{ji}(\bp_i,\bq_i),\bp'_j ) \,\Psi (\bp'_j ,\pmb{\mathcal{Q}}_{ji}(\bp_i,\bq_i))
\cr
&& + \int d^2p'_k\, V_k (\pmb{\mathcal{P}}_{ki}(\bp_i,\bq_i),\bp'_k )\, \Psi (\bp'_k ,\pmb{\mathcal{Q}}_{ki}(\bp_i,\bq_i)). \quad
\label{Sch_test}
\end{eqnarray}
By using the coordinate system defined in Fig.~\ref{fig_coordinate_1}, Eq.~\eqref{Sch_test} can be written as 
\begin{eqnarray}
&& E_t\Psi(p_i,q_i,\phi_i) = 
\left[ \dfrac{p_i^2}{2\mu_{jk}} + \dfrac{q_i^2}{2\mu_{i,jk}} \right] \,\Psi(p_i,q_i,\phi_i)
\cr
&& +\int_0^{\infty} dp'_i \, p'_i\int_0^{2\pi}d\phi'_i \,V_i(p_i,p'_i,\phi'_i)\,\Psi(p'_i,q_i,\phi_i - \phi'_i)
\cr
&& + \int_0^{\infty} dp'_j \, p'_j \int_0^{2\pi}\,d\phi'_j \, V_j (\mathcal{P}_{ji}(p_i,q_i,\phi_i),p'_j,\phi_{\mathcal{P}_{ji},p'_j} ) 
\cr && \quad \times\Psi (p'_j ,\mathcal{Q}_{ji}(p_i,q_i,\phi_i),\phi_{\mathcal{Q}_{ji},p'_j})
\cr
&& + \int_0^{\infty} dp'_k \, p'_k\,\int_0^{2\pi}\,d\phi'_k \, V_k (\mathcal{P}_{ki}(p_i,q_i,\phi_i),p'_k,\phi_{\mathcal{P}_{ki},p'_k} )
\cr && 
 \quad \times \Psi (p'_k ,\mathcal{Q}_{ki}(p_i,q_i,\phi_i),\phi_{\mathcal{Q}_{ki},p'_k}).
\end{eqnarray}
%

\section{Numerical Methods} \label{sec:num}
The coupled Faddeev integral equations \eqref{Faddeev_cpupled_integral} have an eigenvalue equation form of $\lambda \ \psi = {\cal K} (E) \cdot \psi$ with the eigenvalue $\lambda = 1$ and an eigenvector composed of three Faddeev components
$
\psi = \left ( \begin{array}{l}
 \psi_i \\
 \psi_j \\
 \psi_k
 \end{array}\right ).
$
We solve the eigenvalue equation with the Lanczos iterative method, which is successfully implemented in two-, three-, and four-body bound state calculations \cite{MOHSENI2021136773, hadizadeh2020three, PhysRevC.102.044001, PhysRevC.90.054002, PhysRevLett.107.135304, PhysRevC.77.064005, hadizadeh2007four}.
Details of the implementation of this Lanczos technique are discussed in Appendix C2 of Ref.~\cite{PhysRevA.85.023610}.

We start the iteration process with an initial Gaussian guess for Faddeev components and stop it after 10-15 iterations. As the kernel of the eigenvalue equation ${\cal K} (E)$ is energy dependent, the solution of the eigenvalue equation can be started with an initial guess for the 3B binding energy, and the search in the binding energy is stopped when $ \vert \lambda -1 \vert \le 10^{-6}$. To discretize the continuous momentum and angle variables, we use the Gauss-Legendre quadratures with a linear mapping $\phi = \pi (1+x)$ for angle variables and a hyperbolic mapping $p = \frac{1+x}{1-x}$ for the magnitude of Jacobi momenta. 

A typical number of mesh points for angle variables is 60, and for the magnitude of Jacobi momenta is 300. The solution of coupled Faddeev integral equations demands a huge number of 3B interpolations on the Faddeev components $\psi_j(\mathcal{P}_{ji},\mathcal{Q}_{ji}, \phi_{ji} ) $ and $\psi_k(\mathcal{P}_{ki},\mathcal{Q}_{ki},\phi_{ki})$ for shifted momentum and angle variables in each iteration step. We use the Cubic Hermite spline interpolation of Ref.~\cite{huber1997new} for its high computational speed and accuracy. To avoid extrapolations outside the Gauss-Legendre grids, we add an extra point $0$ to all Jacobi momenta grids and two extra points $0$ and $2\pi$ to all angle grids.

\section{3B energy eigenvalues}
\label{App_trion_energies}
In Table \ref{table_trion_l0}, we provide our numerical results for 3B energy eigenvalues obtained from the solution of the coupled Faddeev integral equations \eqref{Faddeev_cpupled_integral}, for the Rytova-Keldysh potential given in Eq.~\eqref{eh_RK}, with two screening schemes for electron-electron interactions shown in Fig.~\ref{fig_RK_screening} and given in Eq.~\eqref{eq:RKregulated}, for different values of screening parameter $l_0$ as a function of the number of mesh points for Jacobi momenta $N_p = N_q$.

\nocite{}

\bibliography{references}

\end{document}